 \documentclass[manuscript]{aastex}

\def\lesssim{\mathrel{\hbox{\rlap{\hbox{\lower4pt\hbox{$\sim$}}}\hbox{$<$}}}}
\def\gtrsim{\mathrel{\hbox{\rlap{\hbox{\lower4pt\hbox{$\sim$}}}\hbox{$>$}}}}

\usepackage{graphicx}
\usepackage{color}
\newcommand{\beq}[1]{\begin{equation} #1 \end{equation}}

\newcommand{\risco}{r_\mathrm{ISCO}}

\newcommand{\rhor}{r_\mathrm{hor}}

\newcommand{\Rmax}{r_\mathrm{max}}
\newcommand{\Rmin}{r_\mathrm{min}}
\newcommand{\rpmax}{r_{p_\mathrm{max}}}
\newcommand{\rin}{r_\mathrm{in}}

\begin{document}

\title{DEPENDENCE OF INNER ACCRETION DISK STRESS ON PARAMETERS:
THE SCHWARZSCHILD CASE}

\author{Scott C. Noble}
\affil{Center for Computational Relativity and Gravitation\\
School of Mathematical Sciences\\
Rochester Institute of Technology\\
78 Lomb Memorial Drive\\
Rochester, NY 14623}

\email{scn@astro.rit.edu}

\and

\author{Julian H. Krolik}
\affil{Physics and Astronomy Department\\
Johns Hopkins University\\ 
Baltimore, MD 21218}

\email{jhk@jhu.edu}

\and

\author{John F. Hawley}
\affil{Department of Astronomy\\
University of Virginia\\
Charlottesville, VA}

\email{jh8h@virginia.edu}

\begin{abstract}

We explore the parameter dependence of inner disk stress in black hole
accretion by contrasting the results of a number of simulations, all
employing 3-d general relativistic MHD in a Schwarzschild spacetime.
Five of these simulations were performed with the intrinsically conservative
code HARM3D, which allows careful regulation of the disk aspect ratio, $H/R$;
our simulations span a range in $H/R$ from 0.06 to 0.17.
We contrast these simulations with two previously reported simulations in a
Schwarzschild spacetime in order to investigate possible dependence
of the inner disk stress on magnetic topology.  In all cases, much care was
devoted to technical issues: ensuring adequate resolution and azimuthal extent,
and averaging only over those time-periods when the accretion flow is in
approximate inflow equilibrium.  We find that the time-averaged radial-dependence
of fluid-frame electromagnetic stress is almost completely independent of
both disk thickness and poloidal magnetic topology.  It rises smoothly inward
at all radii (exhibiting no feature associated with the ISCO) until just
outside the event horizon, where the stress plummets to zero.  Reynolds
stress can also be significant near the ISCO and in the plunging region;
the magnitude of this stress, however, depends on both disk thickness and magnetic
topology.  The two stresses combine to make the net angular momentum accreted
per unit rest-mass 7--$15\%$ less than the angular momentum of the ISCO.

\end{abstract}


\section{Introduction}

At the very beginning of accretion disk studies, their overall
properties were analyzed by applying the constraints of energy
and angular momentum conservation to the simplest reasonable
approximation to their structure: they were assumed to be time-steady
and axisymmetric, and any internal vertical structure was integrated
over \citep{NT73,SS73,1974ApJ...191..499P}.  The equation of energy
conservation can be closed by counting the energy carried by photons
to infinity, but no such ready closure exists for the angular momentum
equation; angular momentum can be conserved for {\it any} rate of angular
momentum transport through the disk provided it does not vary with radius.
Consequently, it was necessary to guess the angular momentum flux in
order to complete the solution.  A convenient way to parametrize this
guess is in terms of the net angular momentum accreted onto the black
hole per accreted rest-mass, $j_{\rm net}$.  The choice made by the
original papers, and still widely-used today, is to suppose that no
stresses act on the flow from the innermost stable circular orbit (ISCO)
inward to the event horizon; if so, $j_{\rm net} = u_\phi (\rm{ISCO})$, the
orbital angular momentum of a test-particle at the last stable orbit
(here $u_\mu$ is the covariant four-velocity).

However, this was never more than a heuristic guess.  As remarked by
\cite{T74}, although the zero-stress boundary condition is plausibly
motivated by hydrodynamic reasoning---the inertia of matter inside the
ISCO should always be much less than that in the stable-orbit portion of
the disk outside the ISCO---it might well be invalid if magnetic fields
are important.  In fact, one of the crucial things we have learned in the
years since the 1970s is that magnetic fields are, in fact, essential
to accretion due to the presence of the magnetorotational instability
(MRI) \citep{BH98}.  On that basis, this traditional boundary
condition has been questioned \citep{K99,G99}, and a number of numerical
simulations of global disks \citep{HK01,AR01b,HK02,MM03,GSM04,KHH05} have
demonstrated that magnetic stresses near the ISCO and in the plunging
region can be sizable.

On the other hand, it has also been suggested that the magnitude of these
inner-disk stresses may be a function of disk parameters, notably its
thickness.  This was the result, for example, of an argument based on a
hydrodynamic model with constant sound speed \citep{AP03}.  Parameterizing
the disk thickness in terms of the ratio of its density scale height $H$
to radial position $r$, \citet{RF08} found that the plunging region stress
in a pseudo-Newtonian MHD simulation with $H/R =0.05$ was rather smaller
than in the analogous simulations of \citet{HK01,HK02} in which $H/R$
was 2--3 times larger\footnote{We also use $R$ to represent the 
radial coordinate, i.e. $r=R$.}.  Similarly, \citet{Shafee08} found that for a disk
with $H/R \simeq 0.06$--0.08 simulated with 3-d MHD in a Schwarzschild
metric, the stress in the plunging region was significantly smaller than
had been found in other simulations with aspect ratios a few times larger
computed with different codes and somewhat different physical assumptions.

It is the goal of this paper to explore how the inner disk stress depends
on parameters, particularly disk thickness, but also magnetic geometry.
To test the former dependence, we have performed a new series of
fully general relativistic 3-d MHD simulations with aspect ratios $H/R
\simeq 0.06$, $\simeq 0.10$, and $\simeq 0.17$, all computed with the
same code in the Schwarzschild metric and using appropriately scaled
initial conditions.  To explore the latter, we review results from
previous relativistic disk simulations and, in particular, make
detailed use of data from two previously reported simulations using 
a Schwarzschild spacetime.  The disks in these two simulations
had almost identical thickness, but in one case the initial magnetic
field was a set of nested dipolar loops, while in the other the initial
field was entirely vertical.  Before presenting the results of these
simulations, we will also discuss the importance of a number of technical
considerations---particularly having to do with spatial resolution and
the establishment of inflow equilibrium---to obtaining meaningful results.

\section{Simulation Details}

The new simulations reported here were made using the code HARM3D,
an intrinsically conservative method to solve the equations of 3-d MHD
in an arbitrary metric.  This new code is described in \cite{Noble09};
see also \cite{GMT03} for a description of the earlier axisymmetric
version HARM, and \cite{Noble06} for additional details on the primitive variable
solver.  We employ the same methodology as before, with only a few exceptions.
In the following summary of this code's techniques, we emphasize those
points that are different from the previous description or particular to
the simulations discussed in this paper.

One of the principal aims of the present work is to study the influence
of disk thickness $H$ on the stress at the ISCO.  We define it
as the density moment in the coordinate frame:
\begin{equation}
H \equiv \int \, d\theta d\phi \sqrt{-g} \, \rho \sqrt{g_{\theta\theta}}
     |\theta - \pi/2| \, / \int \, d\theta d\phi \, \sqrt{-g} \rho \quad , 
\label{scaleheight}
\end{equation}
where $g_{\mu\nu}$ is the metric, $g$ is the determinant of the metric, and 
$\rho$ is the rest-mass density.  When the density profile follows a Gaussian 
distribution with standard deviation $H_G$,  $H = \sqrt{2/\pi}H_G = 0.798 H_G$.
As in \cite{Noble09}, we regulate the thickness by cooling bound portions of 
the disk when the local temperature is greater than some 
target temperature $T_*(r)$.  In terms of intensive quantities, 
bound matter satisfies $\left( \rho + u + P \right)u_t > - \rho$
and gas has temperature above the target when
$\left(\Gamma - 1\right) u / \rho > T_*$; here, $P$ is the gas pressure, 
$u$ is the internal energy density, $u_\mu$ is the  fluid's 4-velocity, 
and $\Gamma$ is the adiabatic index of the equation of state:  
$P = \left(\Gamma - 1\right) u$.
The relativistic enthalpy $h \equiv 1 + (u+P)/\rho$.
We set $\Gamma = 5/3$ throughout.
The optically-thin cooling function is implemented by modifying 
the stress-energy conservation equation
to include a sink term: $\nabla_\mu T^{\mu\nu} =
-{\cal L}u^\nu$.

The fluid-frame emissivity, ${\cal L}$, and $T_*$ are
designed so as to keep the density scale height at the desired value.  
The emissivity is the same as before but we slightly modified
$T_*$ to include a neglected relativistic correction. The new target temperature is 
\beq{
T_* = \frac{\pi}{2} \frac{R_z(r)}{r}  \left[ \frac{H(r)}{r} \right]^2 
\quad , \label{new-target-temperature}
}
where $R_z$ is the relativistic correction to the vertical component of gravity 
\citep{alp97,k99book}\footnote{Note that Equation~(\ref{Rz-corrected}) corrects 
Equation~(7.43) of \cite{k99book}, which propagated a typographical error in
\cite{alp97}.}:
\beq{ 
R_z(r) = \frac{1}{r} \left[ l_k^2 - a^2 \left(\epsilon_k^2 - 1 \right) \right]  
\quad , \label{Rz-corrected}
}
Here, $l_k$ and $\epsilon_k$ are the specific angular momentum ($u_\phi$) 
and energy ($u_t$) of circular time-like geodesics in the equator of a 
black hole with spin parameter $a$.  For $r < \risco$, $l_k$ and $\epsilon_k$
remain at their ISCO values. 

All of the new simulations were performed in a Schwarzschild spacetime ($a = 0$)
described in terms of Kerr-Schild coordinates and run for durations of $12000M$
to $15000M$ (in our units, $G=c=1$, so that both time and distance
have units $M$, the mass of the central black hole).  In all cases, the
initial condition was a hydrostatic torus, but we examined two varieties
of this state: in one the radial coordinate of the pressure maximum $\rpmax =35M$
and the inner edge $\rin=20M$; in the other, $\rpmax=25M$
and $\rin=15M$.  The former set of parameters were
chosen to match those of \cite{Shafee08}, the latter to match those
of \cite{2003ApJ...599.1238D}.  In the remainder of this paper, we
will refer to the former set as ``HR" and to the latter set as
``LR" because the grid schemes used for the former were in general higher resolution
than for the latter.  The $q$ parameter determines the angular velocity
profile in the initial torus ($\Omega \propto r^{-q}$).  The choice
of $q$ along with the choice of pressure maximum and inner torus edge
determines the characteristic vertical thickness of the initial torus.
For LR simulations, $q$ was set to $1.68$, while for HR, $q$
was determined by requiring the initial disk's thickness at $r=\rpmax$ to be equal 
to the run's target thickness.   To study the effect
of different disk thicknesses, we ran HR simulations for three
different target temperatures, chosen to make $H/R\simeq 0.05$, $0.08$, and
$\simeq 0.16$; these were designated ``Thin'', ``Medium'', and ``Thick'',
respectively.  We ran only LR simulations for the Thin and Medium cases.  Finally,
in all five simulations,
the initial magnetic field consisted of dipole poloidal loops, with
field lines following density contours, and with amplitude set such that 
the mean initial plasma $\beta=100$.  Turbulence was seeded by adding random 
perturbations to $u$ at the $1\%$ level. 

Boundary conditions were imposed through assignment of primitive variables 
in ghost zones; the primitive variables are $\rho$, $u$, $v^i$
(spatial velocity components), and $B^i$
(magnetic field components).  The $v^i$ are spatial components of the
four-velocity as seen by observers in the ZAMO frame; the $B^i$ are the spatial
components of the magnetic field as represented in the Maxwell field tensor, i.e.,
$B^i \equiv {^{^*}\!\!F}^{it}/\sqrt{4\pi}$.
Outflow boundary conditions were taken both at $r=\Rmax$ and $r=\Rmin$:
all primitive variables are extrapolated at $0^\mathrm{th}$-order into the ghost zones,
but $u^r$ is set to zero---and $v^i$ recalculated---whenever
it points into the domain.  In order to prevent numerical boundary effects from
propagating outside the trapped surface, we chose $\Rmin$ so that the numerical
domain extended $5$ to $25$ cells inside the event 
horizon.  When there was a cutout around the polar axis (see grid details
below), reflective boundary conditions
are imposed on the perpendicular vector components while all other quantities 
are extrapolated beyond the cutout with $0^\mathrm{th}$-order accuracy. 
If the cutout size is negligible, ghost zone values are set so as to make
the variables continuous across the pole.  In order to gain a factor of four in
computer resources, we simulated only a quarter of the azimuthal domain, employing
periodic boundary conditions linking $\phi = 0$ and $\phi = \pi/2$.

Adequate resolution throughout the accretion flow and throughout the duration of
the simulation is vital to ensure quantitative accuracy, particularly
for magnetic effects.  To explore the consequences of resolution
effects, we used two different grid schemes, one for HR, the
other for LR.  Both were designed with an eye toward satisfying
several conflicting criteria.  On the one hand, it is always desirable
to have the finest feasible spatial resolution.  Toward that end, the grid
scheme should provide at
least several dozen cells per scale height on either side of the
equatorial plane, and the poloidal cell aspect ratio should never be
too large.  In addition, the number of cells within a wavelength of the
fastest-growing MRI mode should only occasionally fall
below $\simeq 6$, the minimum resolution level at which the mode grows
at the correct rate \citep{Sano04}.  On the other hand, to fit within
existing computational resources, the total cell count cannot be too
great nor the time step too small.  

Each run used a unique discretization tailored to its particular 
target thickness.  HARM3D accomplishes nonuniform discretization through 
continuous coordinate transformations from an underlying uniform mesh.
The finite volume equations are discretized with respect to points uniformly
distributed in coordinates, $x^\mu$, which are placed nonuniformly in the spherical 
coordinate system $r$, $\theta$, and $\phi$.  The center of cell $C_{ijk}$ is 
located at $(x^1_{i+1/2}, x^2_{j+1/2}, x^3_{k+1/2})$, where 
$x^\mu_i \equiv x^\mu_0 + i \Delta x^\mu$  and $\Delta x^\mu$ is a cell's extent 
in the $x^\mu$ direction.  We choose  $t=x^0$, $\phi=x^3$, 
and 
\beq{
r_i = e^{x^1_i}  \quad , \label{r-to-x1} 
}
so that $\Delta r/r$ is the same everywhere.  For $N_1$ grid cells along
the $x^1$ axis and minimum and maximum radii $\Rmin$ and $\Rmax$, $\Delta x^1 =
\frac{1}{N_1} \log\left(\frac{\Rmax}{\Rmin}\right)$.

The relationship between $x^2$ and $\theta$ is determined differently
in the HR and LR simulations.  In the LR group, we follow
previous work \citep[e.g.,][]{GMT03,2007CQGra..24..259N,Noble09},
but introduce a ``cutout'' or excised region around the polar axis:
\beq{
\theta(x^2)  \, = \, \theta_c \, + \, \left( \pi - 2 \theta_c \right) x^2 \, 
+ \, \xi \, \sin\left[2\left(\theta_0 + s x^2 \right)\right]
\quad ,  \label{theta-x2-1-def} 
}
where $\theta_c$ is the approximate size of the cutout,
$\theta_0$ and $s$ control the nonlinearity of the transformation, 
and $\xi$ is the amplitude of the nonlinear part.  One drawback 
to this transformation is that if one tries to place a majority of the points 
within the first two scale heights from the equator, the minimum
$\Delta\theta$, which is always found at $\theta = \pi/2$, is so small
that the time step is prohibitively small.

In the HR simulations, we follow the method of \cite{Shafee08}, 
adapted here to include a cutout.  In this method, the sinusoidal nonlinear 
term is replaced with a polynomial:
\beq{
\theta(x^2) = 
\frac{\pi}{2} \left[  1  + \left(1-\xi \right) \left(2 x^2 - 1 \right) 
+ \left( \xi - \frac{2 \theta_c}{\pi} \right) \left( 2 x^2 - 1 \right)^n \right] 
\quad ,  \label{theta-x2-2-def}
}
where $n$ is a positive odd integer, $\theta_c$ is the size of the 
excised region, and $\xi$ is still the relative 
amplitude of the nonlinear term.  Note that near the equator, where 
most of the points are located, the linear term dominates and 
$\Delta\theta(x^2)$ is nearly uniform.  Periodicity of 
$\theta \in [0,2\pi]$ is ensured by making $x^2$ a periodic 
triangle function over $[0,2]$. 
We set $n=9$ whenever equation~(\ref{theta-x2-2-def})
is used for the runs presented here.

Because gridscale dissipation scales with the ratio of cell dimension
to the length scale on which physical quantities vary, cells that are far
from cubical may have effective dissipation properties
that are anisotropic.  This might produce unphysical results because physical
dissipation mechanisms are unlikely to have this property.  For this reason,
we strive to limit the degree of anisotropy in our cell shapes, although cells
longer by factors of a few in the $\phi$-coordinate than in the others are acceptable
because orbital shear tends to draw out features in the azimuthal direction.
At the same time, it is important, especially for small $H/R$ simulations,
to make the $\theta$-direction cell thickness small enough to put an adequate
number of cells within a vertical scale height.  Achieving these conflicting
goals is easier with the HR grid scheme than with the LR scheme
(as illustrated by the data in Table~\ref{tab:simdefs}.
Our HR runs use grids with $\Delta r : r \Delta \theta \simeq  2 : 1$ while
achieving at least $60$ cells per scale height.  In LR simulations, although
the aspect ratios are acceptable at an altitude $\sim H$ away from the
equatorial plane, they are rather extreme close to the plane, and---of these runs---only 
ThinLR has a comparable number of cells per vertical scale height to HR runs 
(see Table~\ref{tab:simresults}).

The time step $\Delta t = \Delta x^0$ is set equal 
to $0.8$ times the shortest cell crossing time for the fastest MHD 
characteristic from the previous time step.

The defining parameters of the simulations are collected in
Table~\ref{tab:simdefs}.  In the first column we state the names 
of the runs.  The body of the name corresponds to the disk
thickness, the suffix refers to the grid resolution.  The target aspect
ratio $H/R$ is shown in the second column.  Note that the ``radius"
to which the scale height is compared is the radial coordinate, which
is not identical to a length in any frame of reference.  Since we
regulate the temperature, not $H$ directly, the actual value will
not in general exactly coincide with the target.
The observed value of $H/R$ is given in Table~\ref{tab:simresults}
and is obtained by averaging over the time interval shown in its  last column
(whose origin is discussed in \S~\ref{sec:equil}) and radially-averaging
from the ISCO ($r =6M$) to the run's $\rpmax$.  The third column, cell-count, is
self-explanatory.  Columns 4 to 8 define the poloidal discretization.
Columns 9 and 10 state the typical cell aspect ratio at the equator and at
two scale heights from the equator, and Column~11 lists the size of the smallest
poloidal extent---$\Delta \theta$---of a cell, which always occurs at $\theta=\pi/2$.

We will also analyze two older simulations, both done with the general
relativistic 3-d MHD code GRMHD (originally described in \cite{dVH03}).
Both began with the gas in a hydrostatic torus.  The first, called
``KD0c,'' was described in \cite{HK06} and had initial conditions
similar to the LR models presented here.  The other, called ``VD0"
and analyzed in \cite{BHK09}, began with a constant intensity vertical
magnetic field filling the annulus $35M \leq r \leq 55M$ and running
through an initial torus with a pressure maximum at $r=40M$.  Unlike the
HARM3D simulations there was no explicit cooling function included.
Rather, the internal energy equation was evolved and the only heating
included was shock heating captured by an artificial viscosity term.
The resulting $H/R$ values are given in Table~\ref{tab:simresults} and,
in terms of the descriptive terms used here, qualify these simulations as
``thick.''

\section{Quality and Consistency Checks}\label{sec:quality}

We are investigating the stress at the ISCO in a steady state accretion
disk with a focus on the role of vertical scale height.  To model the
accretion disk system, we use the equations of conservative ideal MHD
along with an \textit{ad hoc} cooling function to control the disk's thickness.
We can, however, run only a discrete set of simulations, evolved for
a limited time from particular initial conditions, using a grid with
restricted spatial extent and modest resolution.  The conclusions we
obtain must be assessed within the context of the limitations of those
simulations.

In this section we consider the effects of some of these numerical
limitations by developing several quantitative diagnostics to measure
their possible significance.  We begin by considering the adequacy of
the grid resolution,  and then look at how closely the inner disk
approximates a statistical steady state.
Finally, since this study focuses on the
effect of disk scale height for the ISCO stress, we check to see how
well we are able to control that variable.

\subsection{Resolution}
\label{subsec:resolution}

Inadequate resolution can cause a number of numerical artifacts.
For example, the growth rate of the underlying MRI can be suppressed
if there are fewer than $\simeq 6$ zones within a wavelength of the
fastest growing mode \citep{Sano04}.  The MRI produces turbulence,
but only a small range in wavenumber space can be captured, possibly
distorting the properties of that turbulence.   The rate
at which nonlinear mode-mode couplings transfer energy from large-scale
motions to small may be altered.  And resolution that is too coarse may
drive an artificially large rate of magnetic numerical dissipation.

Unfortunately, there is no {\it a priori} standard by which we can
measure whether a given grid scheme either exhibits excessive magnetic
reconnection or improperly evaluated nonlinear mode couplings.
Only through a numerical convergence test, in which increasingly
better-resolved simulations give quantitatively consistent results,
can one demonstrate that resolution artifacts are not influencing
the outcomes.  As a practical matter, however, one carries
out simulations such as these at the highest feasible resolution.
Lower resolution simulations might provide some information, but with the
computational resources at our disposal it has not been possible to improve
sufficiently upon the resolution used in the production simulations to
establish formal convergence.

It is possible, however, to check whether the grid resolution satisfies
certain physical criteria, such as having a sizable number of cells per
vertical scale height and sufficient cells across the fastest growing
MRI wavelength that the linear growth of these modes is correctly described.
Both criteria may be met in the initial conditions, but the data
must be examined throughout the relevant volume (the main disk body) and
for the duration of the simulation to ensure that they continued to be
satisfied.

Data on the mean number of cells per scale height are displayed in
Table~\ref{tab:simresults}.  There we see that in all the HR
simulations, there were at least $\simeq 40$ cells per scale height.
ThinLR was almost as well-resolved as the HR simulations with
respect to this measure, with 30 cells per $H$.  MediumLR has somewhat
fewer cells per scale height, $\simeq 18$, but even with this number
should still be able to resolve well dynamics on the scale of a
fraction of a scale height.

\begin{figure}
\includegraphics[angle=0,scale=0.42]{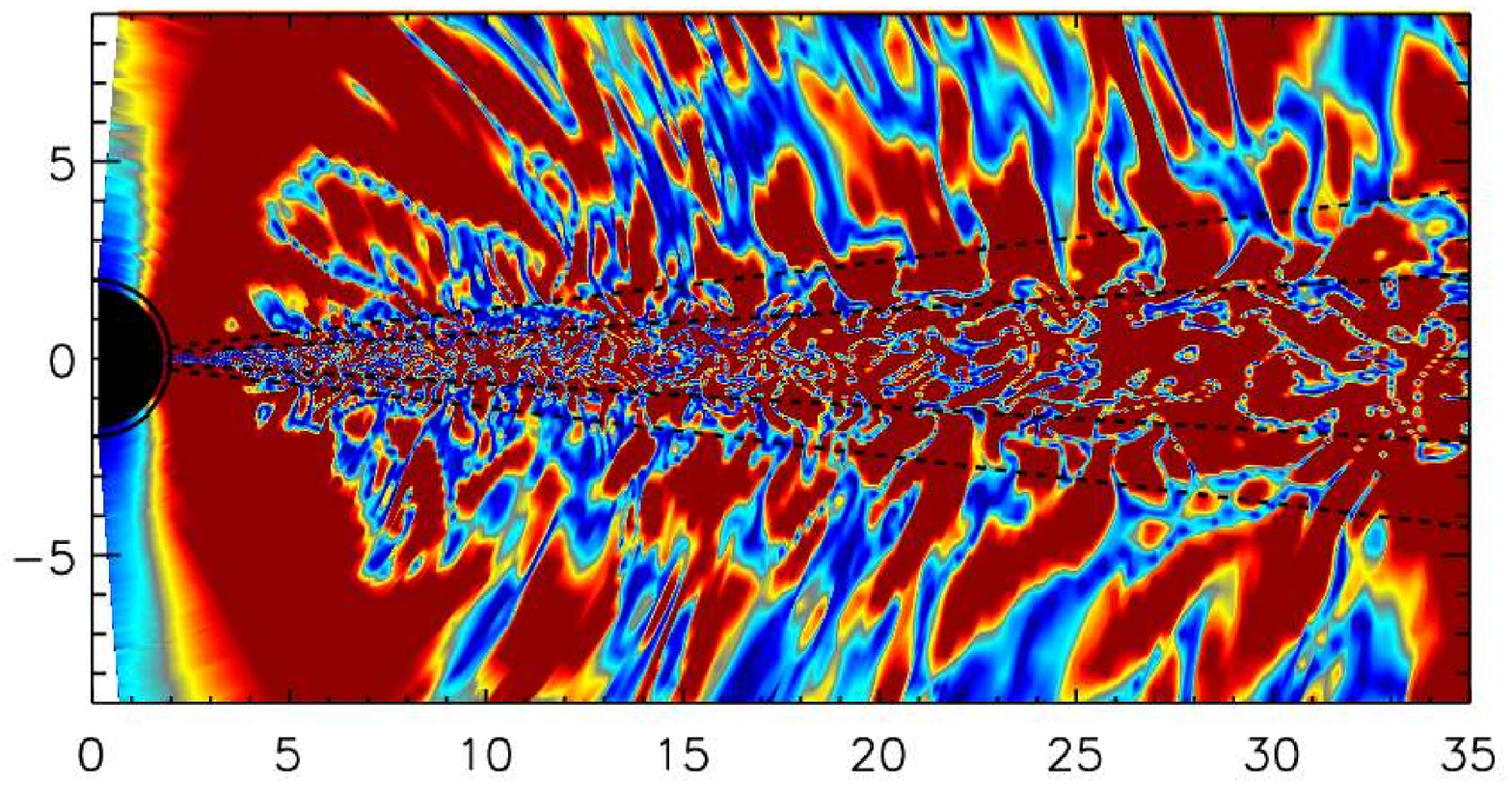}
\includegraphics[angle=0,scale=0.42]{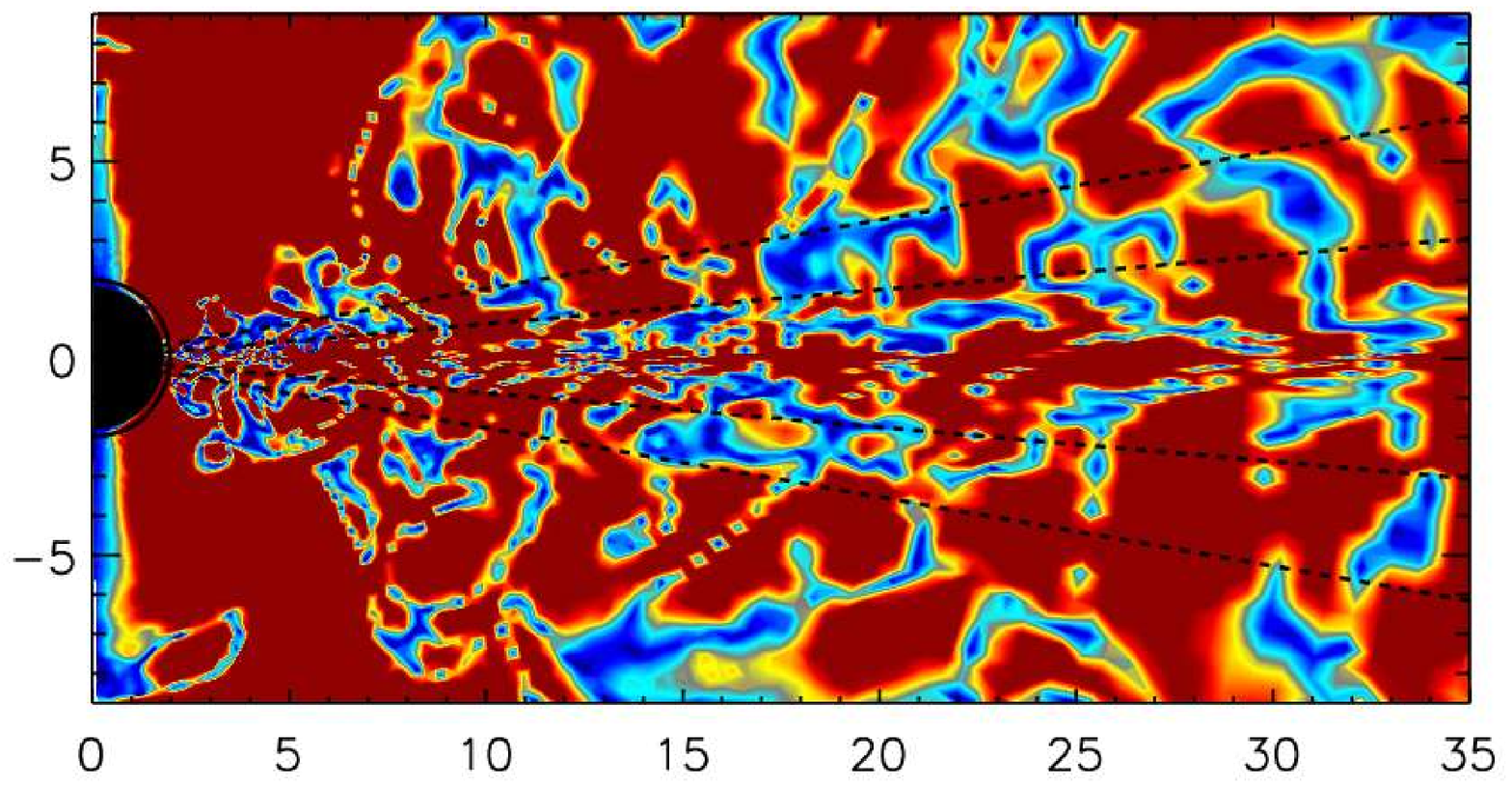}\includegraphics[scale=0.3]{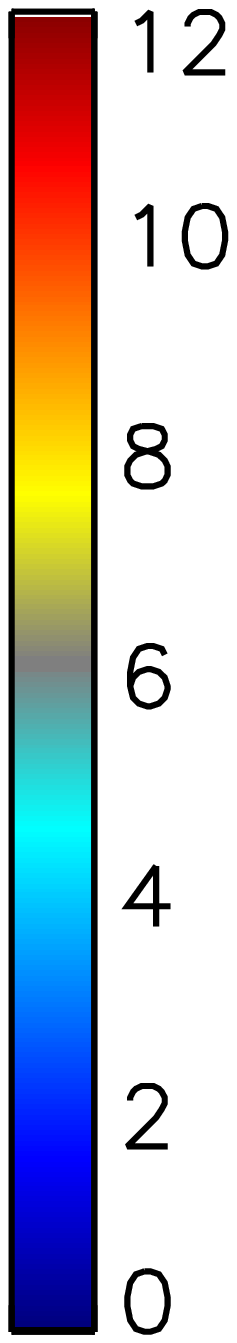}\\
\includegraphics[angle=0,scale=0.42]{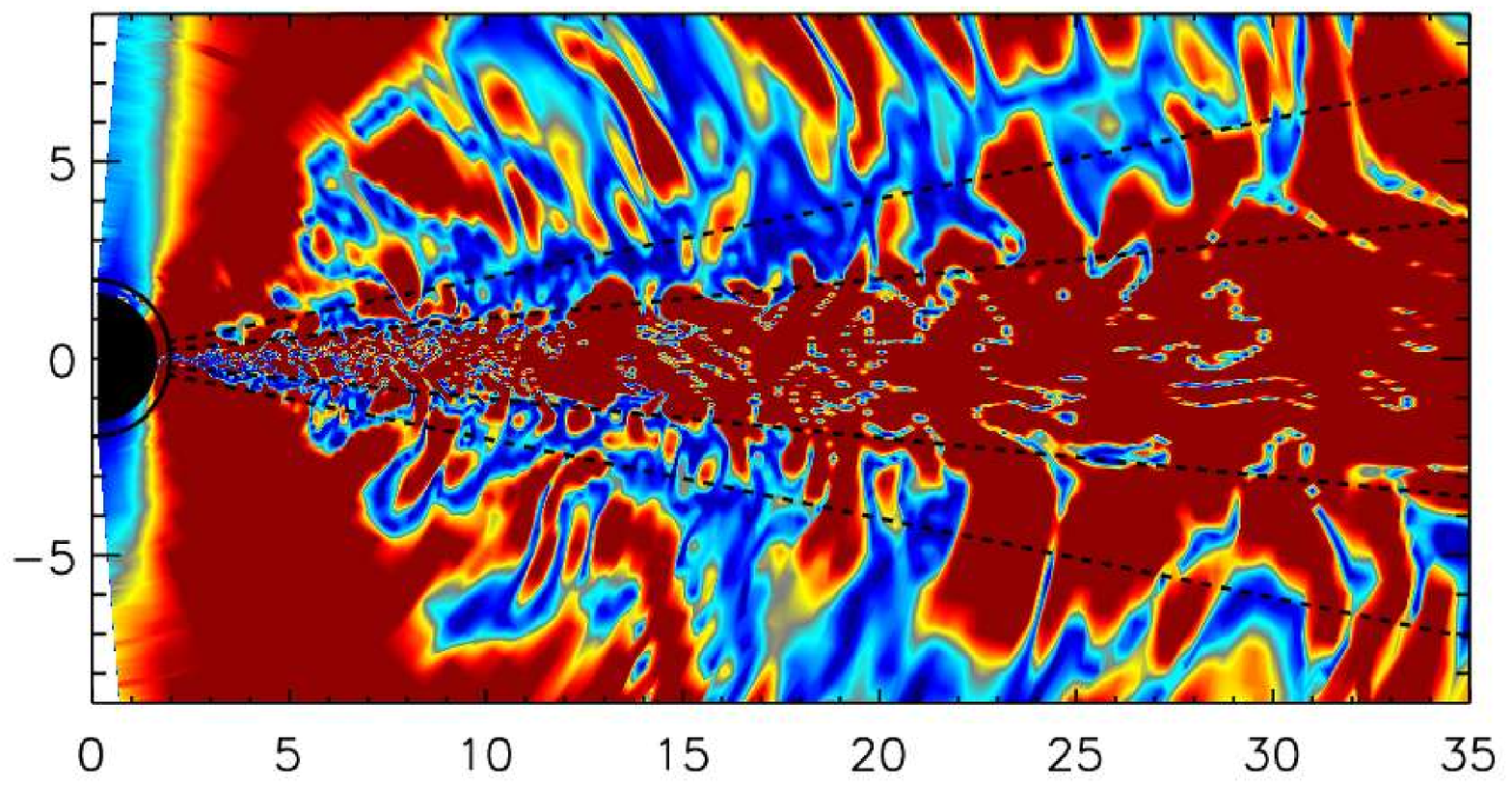}
\includegraphics[angle=0,scale=0.42]{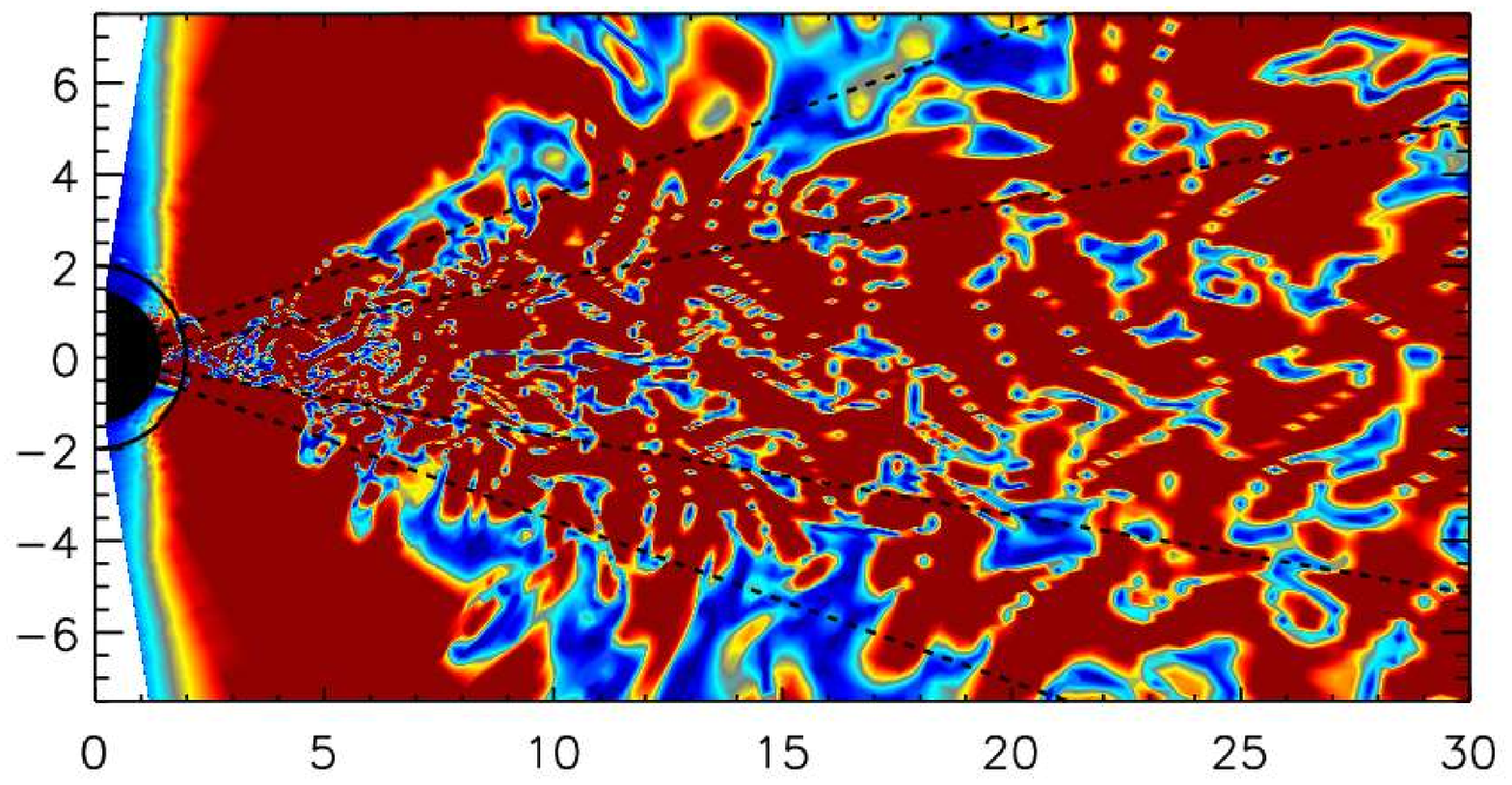}
\caption
{Sample data illustrating resolution of the MRI in simulation ThinHR
(top left), ThinLR (top right), MediumHR (bottom left), and ThickHR
(bottom right).  In each case, the region shown is a poloidal slice
at $\phi=\pi/4$.  Regions with deep red color are well-resolved; those
in blue are poorly-resolved.  The dashed black lines show one and two
scale heights from the midplane.
\label{fig:resolution1}}
\end{figure}

To quantify the quality of resolution of the MRI, we evaluated
the parameter $Q \equiv \lambda_{\rm MRI}/\Delta z$, where both
$\lambda_{\rm MRI}$ and $\Delta z$ are computed in the fluid frame,
\begin{equation}
\lambda_{\rm MRI} \equiv \frac{1}{\sqrt{4\pi \rho} \, \Omega(r)}
            b_\mu \hat e^\mu_{(\theta)} ,
\end{equation}
and
\begin{equation}
\Delta z \equiv dx^\mu \hat e_\mu^{(\theta)},
\end{equation}
where $\hat e^\mu$ and $\hat e_\mu$ are contra- and co-variant tetrad
systems in the local fluid-frame, respectively.  To assure ourselves that
the simulations always satisfied this criterion, we created animations
of $Q$ with frames every $20M$ in time.  In Figure~\ref{fig:resolution1},
we show sample stills from the simulations ThinHR, ThinLR, MediumHR, and
ThickHR, each exhibiting a vertical slice at fixed azimuthal angle.  All of
these simulations exhibited better than adequate resolution
at all times: the typical number of cells across the fastest-growing
mode in the disk body was $\gtrsim 20$.  Azimuthally-averaged versions
of the data shown in Figure~\ref{fig:resolution1} display ratios of
fastest-growing wavelength to cell-size greater than 12 throughout
the entire region within two scale heights of the equatorial plane.
It is important to recognize, however, that even in a superbly-resolved
simulation there will always be the occasional region in which the local
poloidal field strength is small simply because this is a chaotic system
in which fields are free to have either sign.

The importance of $Q$ comes from the requirement that the MRI maintain
turbulence in the face of continual dissipation.  If the MRI growth
rate is artificially reduced due to poor resolution,\footnote{Although
one generally expects poor resolution to limit the strength of the MRI
turbulence, \cite{Fromang07} found that for zero net magnetic flux unstratified
shearing box experiments, increased resolution led to decreased
turbulent stress, appearing to converge to zero as resolution increased.
It now seems that this result arises from another numerical limitation, namely the
lack of any length scale besides that of a grid zone.  Both \cite{Shi09} and
\cite{Davis09} find that convergence to a nonzero stress is
recovered when vertical gravity is added.} and field amplification by MHD
turbulence is too weak, the field strength can become increasingly anemic.
Thus it is possible for a simulation that began with ample cells per
MRI wavelength to suddenly find itself under-resolved.  The end-result
is a field that dies away and a cessation of accretion due entirely to
inadequate resolution.

\begin{figure}
\centerline{\includegraphics[angle=0,scale=0.42]{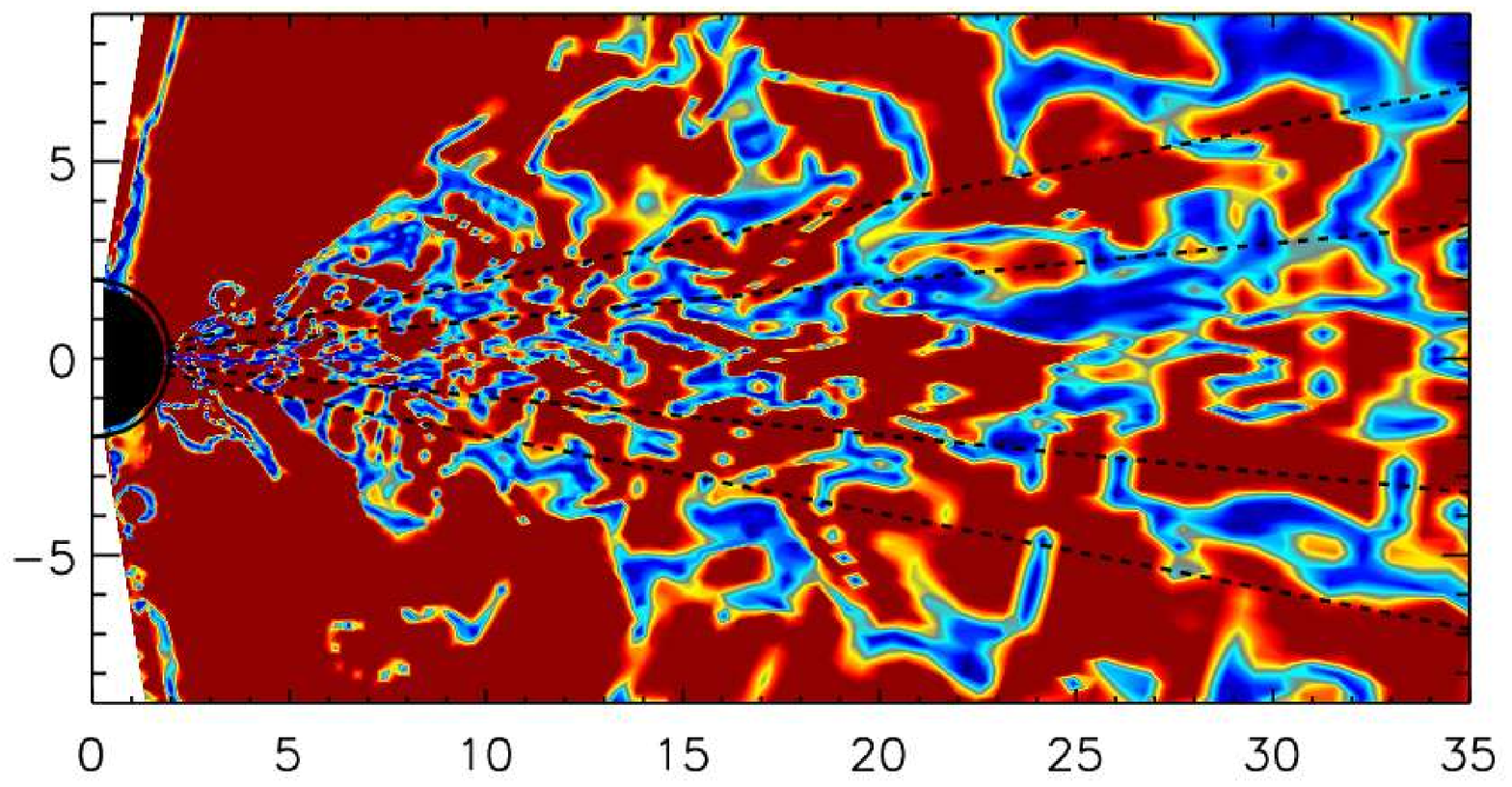}
\includegraphics[angle=0,scale=0.42]{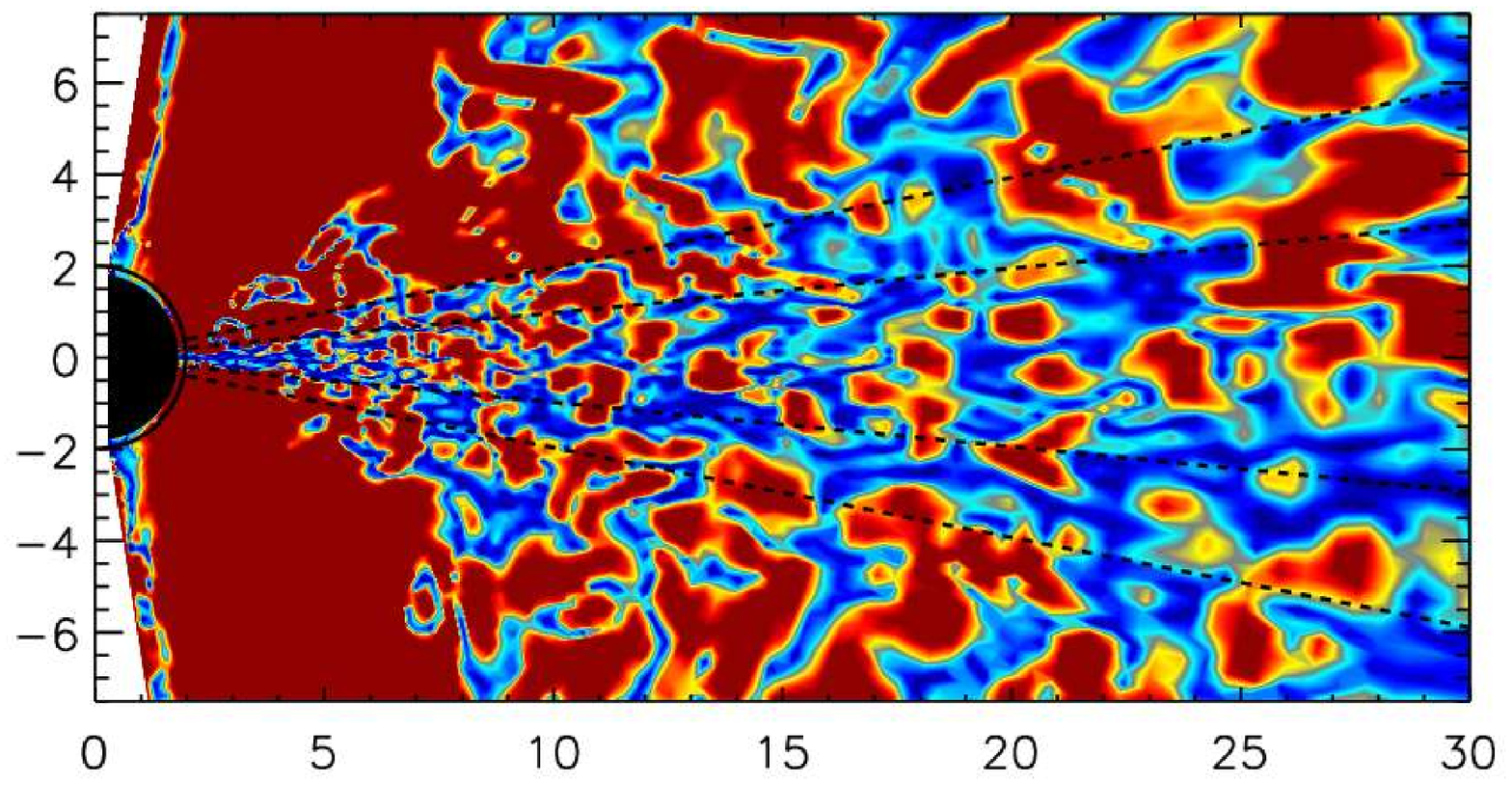}
\includegraphics[scale=0.3]{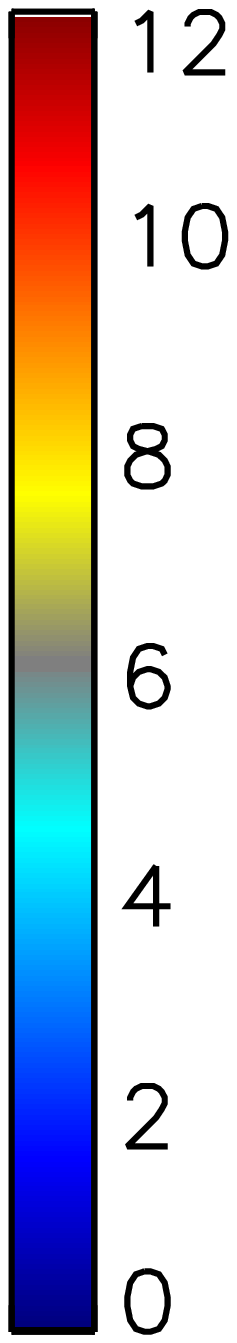}}
\caption
{Sample data illustrating resolution of the MRI in simulation MediumLR
while it was well-resolved (left, at time $t=6000M$) and after it became
poorly-resolved (right, at time $t=12000M$).  In both cases, the region shown
is a poloidal slice.  Regions with deep red color are well-resolved;
those in blue are poorly-resolved.  The dashed black lines show one and
two scale heights from the midplane.
\label{fig:resolution2}}
\end{figure}

This appears to happen in the MediumLR simulation
beginning at $t \simeq 9000M$.  MediumLR demonstrates the 
utility of the resolution diagnostic.  Figure~\ref{fig:resolution2}
shows two stills, at times $6000M$ and $12000M$, from the resolution animation
for this simulation.
Initially well-resolved, MediumLR becomes more poorly resolved as the
field strength diminishes. As a result, the accretion rate also decays.

\begin{figure}
\centerline{\includegraphics[angle=90,scale=0.4]{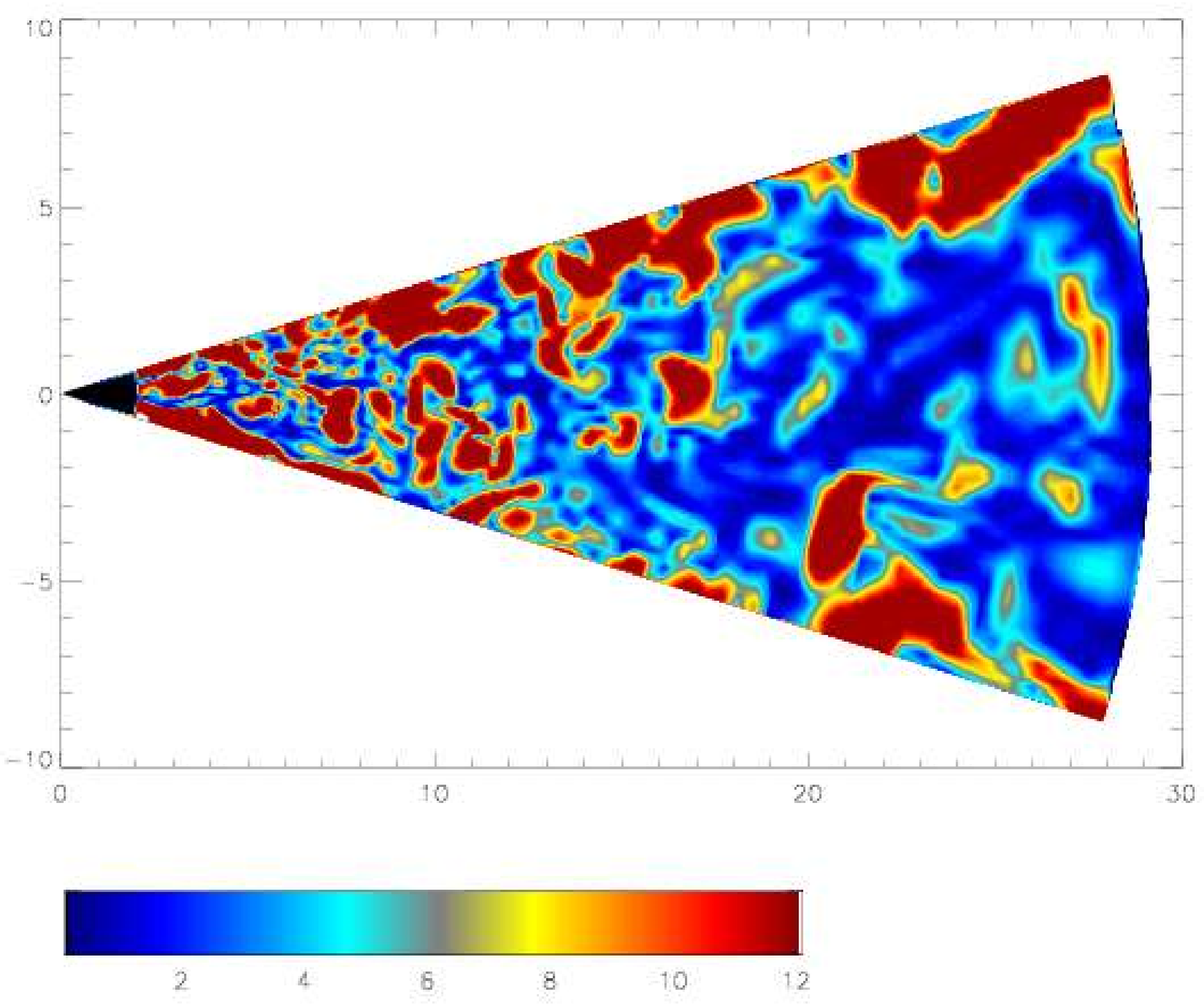}
\includegraphics[angle=90,scale=0.4]{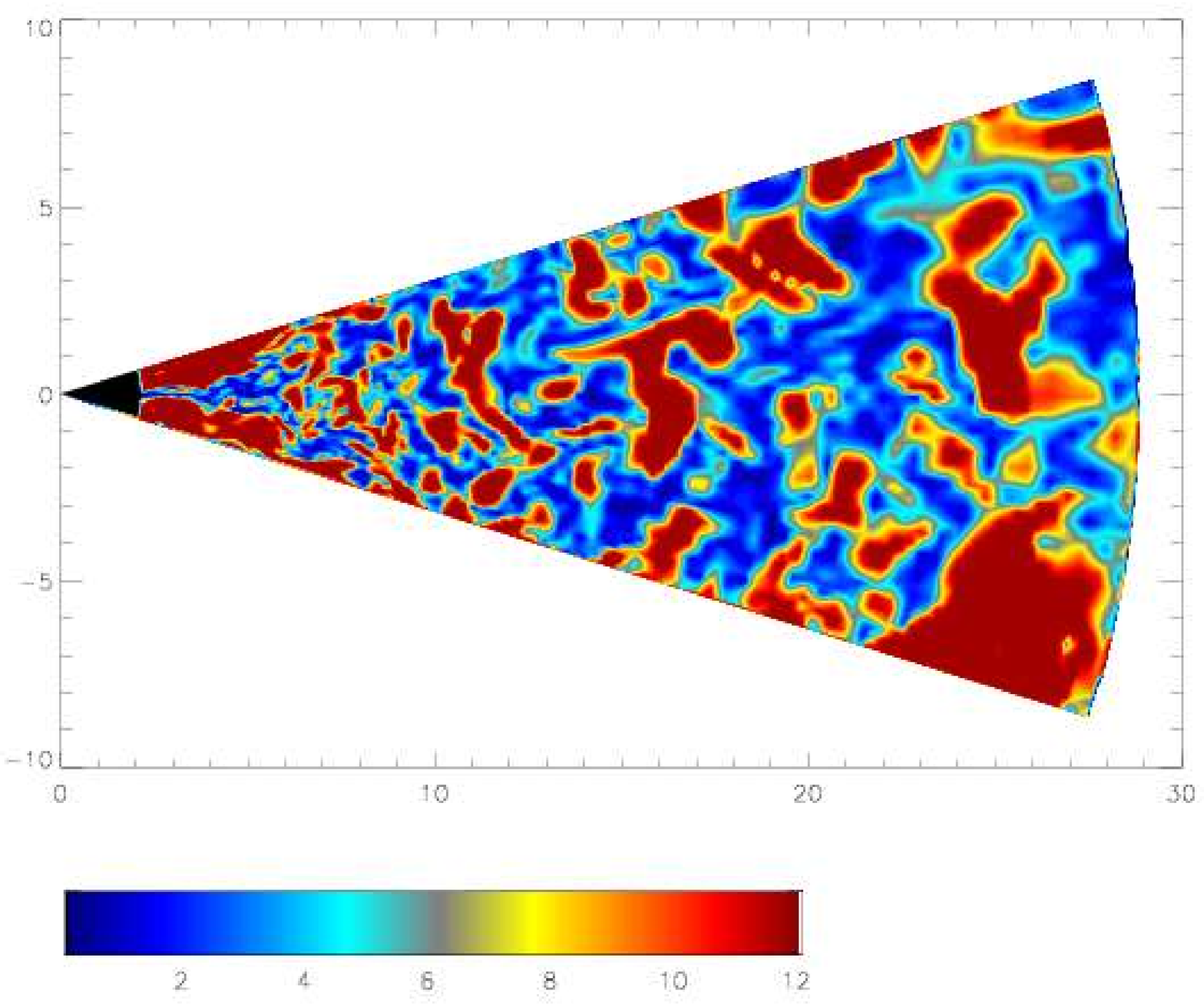}}
\caption
{Sample data illustrating resolution of the MRI in simulation KD0c
at its end-time, $t=10000M$ (left) and in simulation VD0 at its
end-time, $t=20000M$ (right).  In both cases, the region shown
is a poloidal slice extending $0.3$ radians, $\simeq 2H$, from the
midplane.
\label{fig:resolutionGRMHD}}
\end{figure}

The two GRMHD simulations can be analyzed in similar fashion.  By the standards
of the HARM3D simulations, both have relatively few cells per scale height,
only $\simeq 12$--16, not quite as many as in MediumLR.  We also show
sample snapshots of $Q$ in Figure~\ref{fig:resolutionGRMHD}, in each
case showing the final time-step of the simulation.  KD0c was
performed several years ago, and we did not save enough snapshots
to create an animation; of all those available (every $80M$ from $t=8000M$ to
$t=10000M$), the one shown in Figure~\ref{fig:resolutionGRMHD} appears to be the
least well-resolved.  In fact, there is an indication of a secular worsening
of resolution beginning at $t \simeq 9000M$.  In the case of VD0,
we possess more data and have confirmed that at no time was the
resolution substantially poorer than shown here.  As can be seen, in
neither case was the resolution as good as in the best of the HARM3D
simulations, although VD0 at its end-point was better than KD0c at its.

\subsection{Inflow equilibrium}\label{sec:equil}

In the end, we hope to use these evolving simulations to describe
time-steady accretion flows.  Starting from our initial condition, this
state can never be achieved at all radii because the angular momentum
removed from accreting material will be transferred to matter at larger
radius, causing that matter to move outward.  Moreover, because
only a fraction of the initial torus mass is accreted in the duration
of the simulations, the radial surface density profile can relax to the
one associated with inflow equilibrium only within a short distance
outside the initial inner radius of the torus ($20M$ for HR,
$15M$ for LR).

Nonetheless, it is possible to identify a period of time for which
an approximate state of inflow equilibrium really does obtain over a
reasonable dynamic range in radius, subject, of course, to the sorts of
fluctuations that occur in statistically stationary turbulent systems.
To identify that region in both time and space, we impose several tests
relying on the conservation of mass and angular momentum.  
Whenever the disk is in inflow equilibrium, the radial fluxes
of mass and angular momentum should be constant as a function of radius,
but because of turbulent fluctuations, they are constant only in a time-averaged
sense.  The radial range of the equilibrium region is determined
by the range over which the time-averaged values of these quantities
are nearly constant.  To identify the time periods in which equilibrium obtains,
we consider the mass interior to several specific radii $M(<r;t)$ and 
the time-dependence of the specific angular momentum accreted onto 
the black hole, $j_{\rm net}$.
In inflow equilibrium, these should be roughly constant over time.

The mass interior to radius $r$ is defined as the integral of
the mass density over the computational volume from the horizon to radial
coordinate $r$, or 
\beq{
M(<r;t) \equiv \int^r_{\rhor} dr^\prime \, d\theta \, d\phi \, \sqrt{-g} \, \rho 
\quad . 
}
Thus, once inflow equilibrium is established,
the mass within a given radius should stay the same, as the amount of
mass entering from outside is matched by the mass accreted onto the
black hole at the center.  

\begin{figure}
\centerline{\includegraphics[angle=90,scale=0.35]{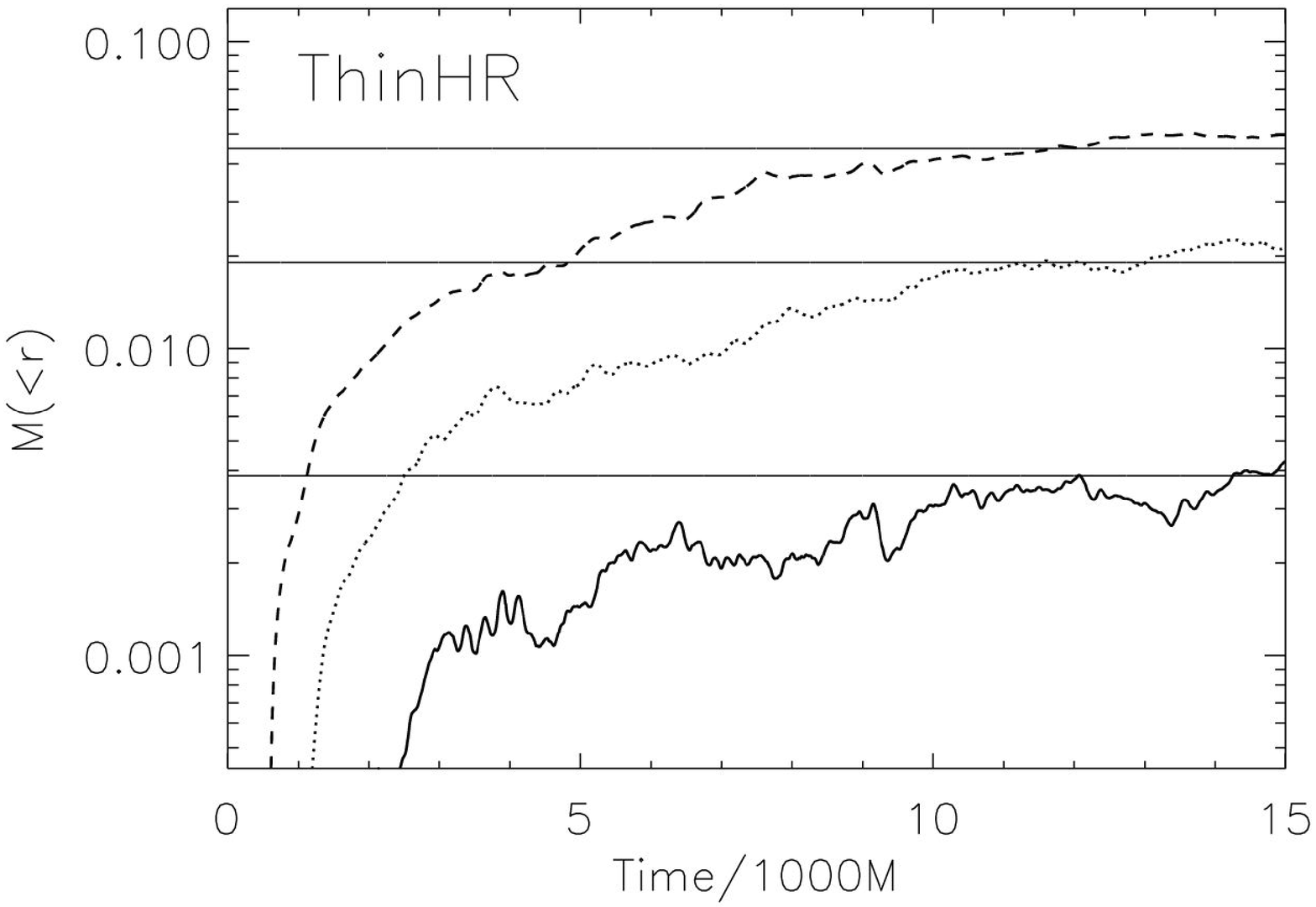}
\includegraphics[angle=90,scale=0.35]{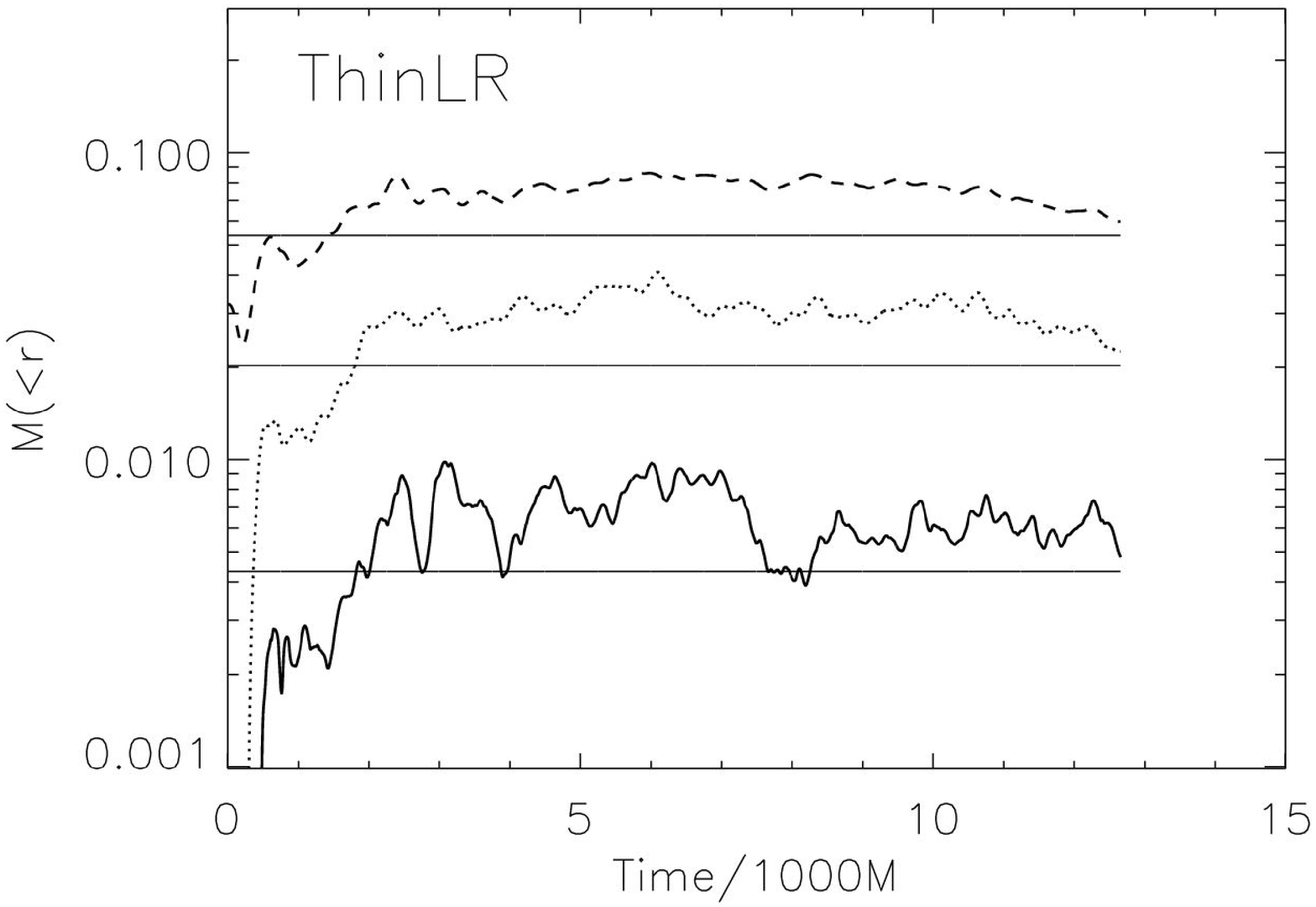}}
\caption
{Mass (normalized to the initial total mass in the disk) inside
$r =10M$ (thick solid curve), $r =15M$ (dotted curve),
and $r =20M$ (dashed curve) for the thin simulations.  The three horizontal
thin solid lines show $90\%$ of the final mass for each of these radii.
(Left) HR.  (Right) LR.
\label{fig:thinmassfillin}}
\end{figure}

\begin{figure}
\centerline{\includegraphics[angle=90,scale=0.35]{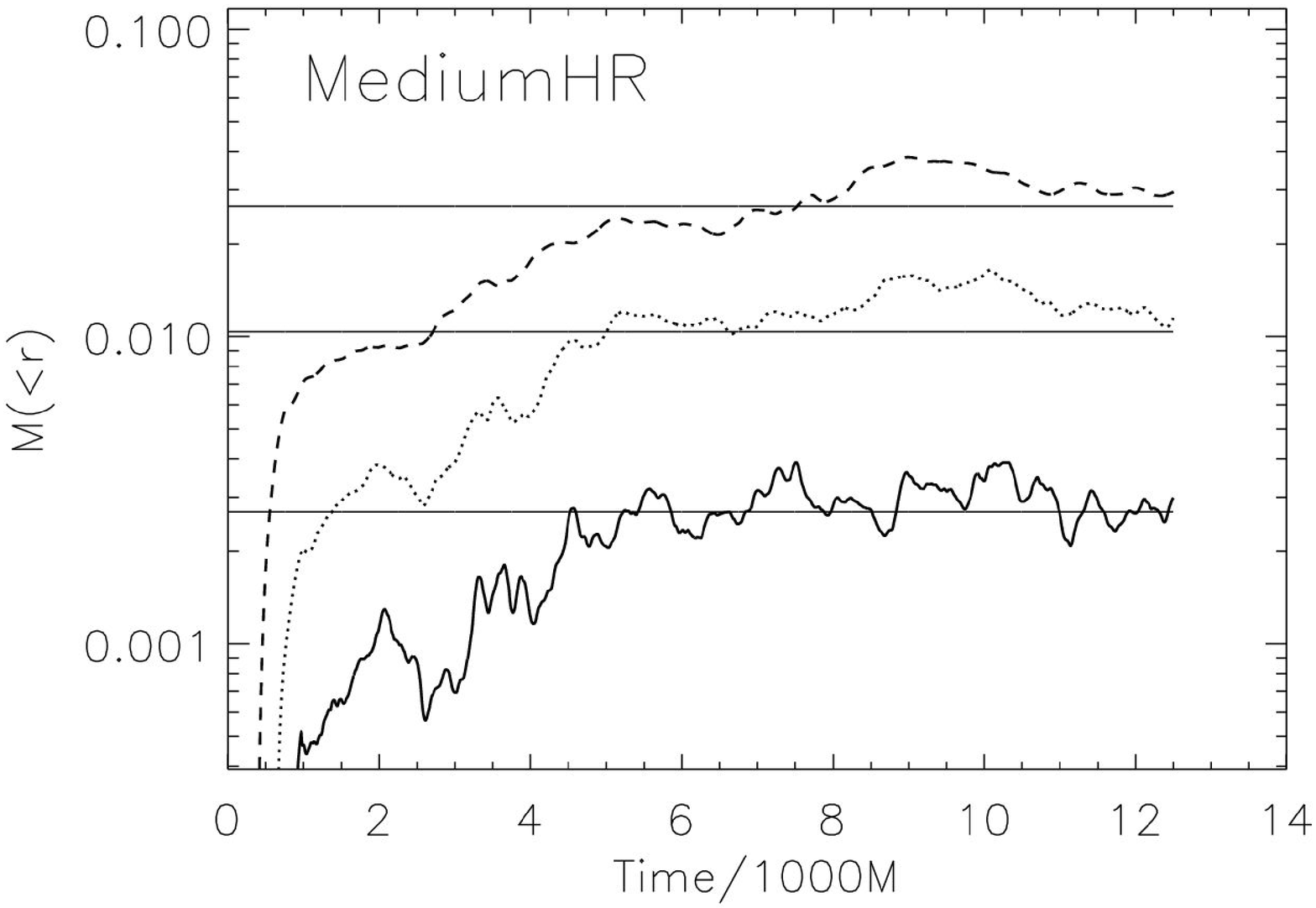}
\includegraphics[angle=90,scale=0.35]{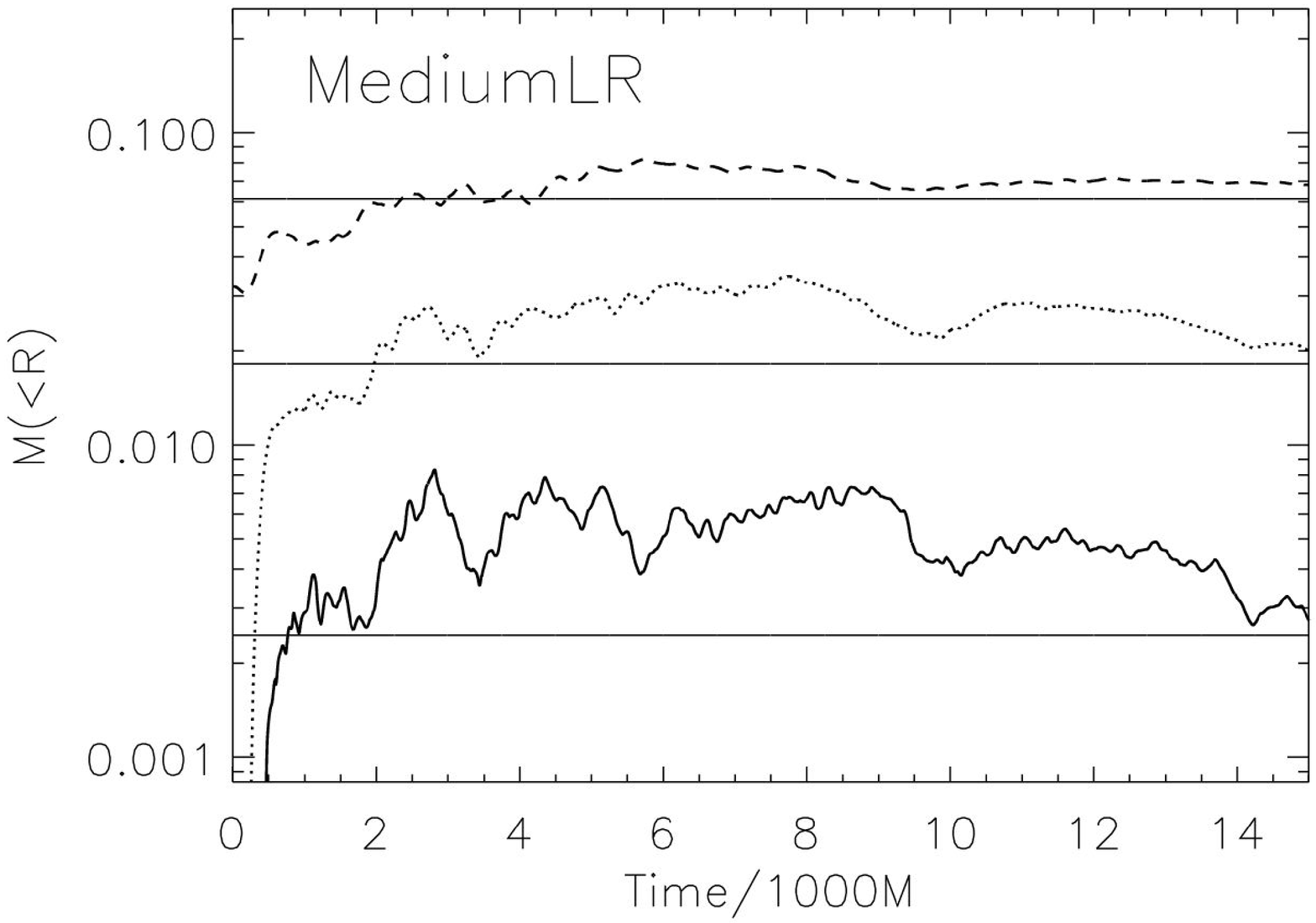}}
\caption{
Mass (normalized to the initial total mass in the disk) inside $r =10M$ (thick solid
curve), $r =15M$ (dotted curve), and
$r =20M$ (dashed curve) for the medium simulations.  The three horizontal
thin solid lines show $90\%$ of the final mass for each of these radii.
(Left) HR.  (Right) LR.
\label{fig:medmassfillin}}
\end{figure}

\begin{figure}
\centerline{\includegraphics[angle=90,scale=0.35]{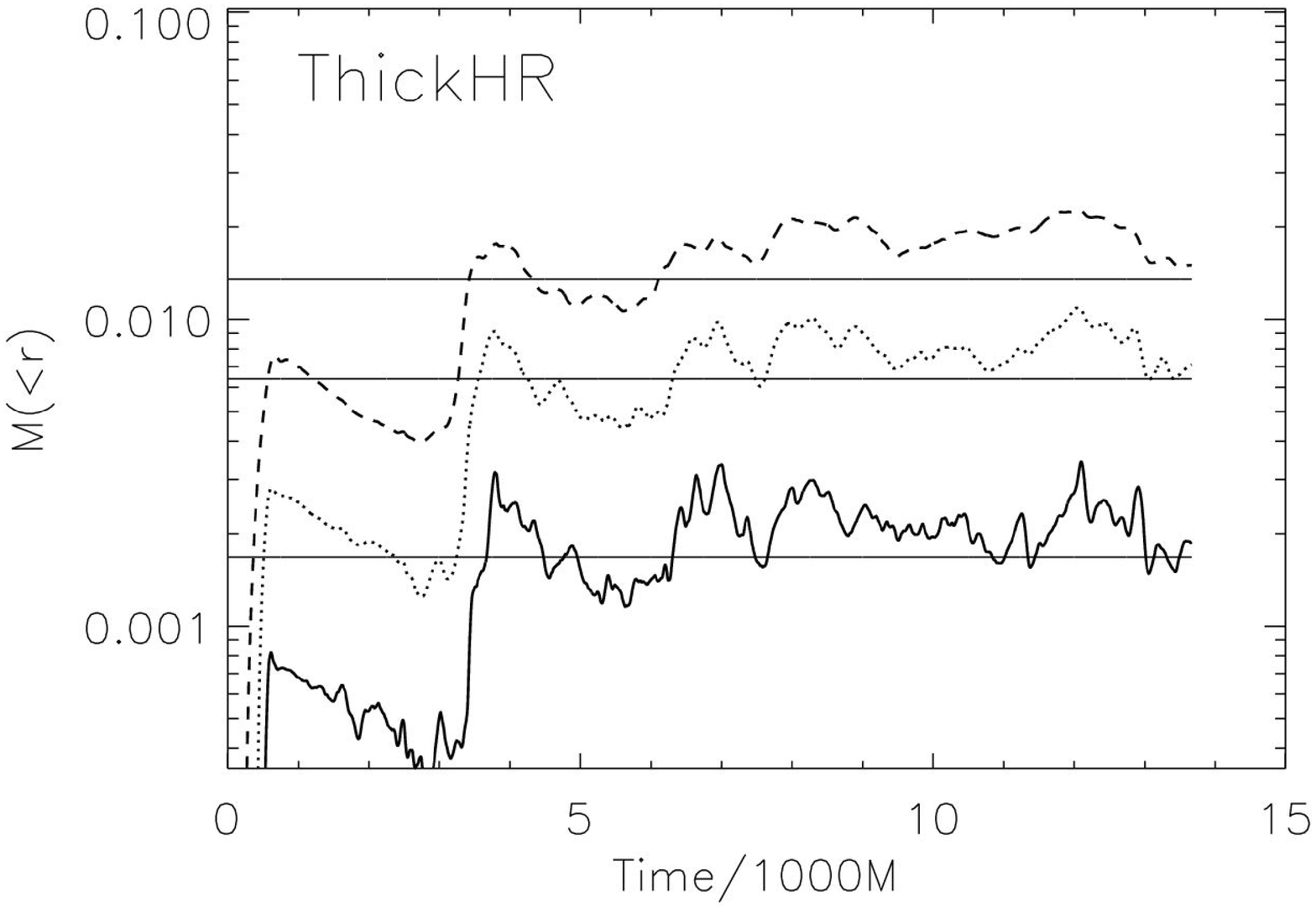}}
\caption{Mass (normalized to the initial total mass in the disk) inside
$r =10M$ (thick solid curve), $r =15M$ (dotted curve),
and $r =20M$ (dashed curve) for the thick simulation.  The three horizontal thin
solid lines show $90\%$ of the final mass for each of these radii.
\label{fig:thickmassfillin}}
\end{figure}

Figures~\ref{fig:thinmassfillin}--\ref{fig:thickmassfillin} display
the history of mass inside $r =10M$, $15M$, and $20M$ for each of the
three disk aspect ratio categories studied.  In all cases,
at early times the mass in the disk grows as accretion from the initial
torus fills in rings at smaller radii.  Eventually, the mass-interior
curves level off, signaling the approach to equilibrium with respect
to this criterion.  In the case of MediumLR, the mass in the inner disk
declines at late times, a symptom of the diminution in the accretion
rate which we attribute to the artificial decay of the magnetic field.

\begin{figure}
\centerline{\includegraphics[angle=90,scale=0.35]{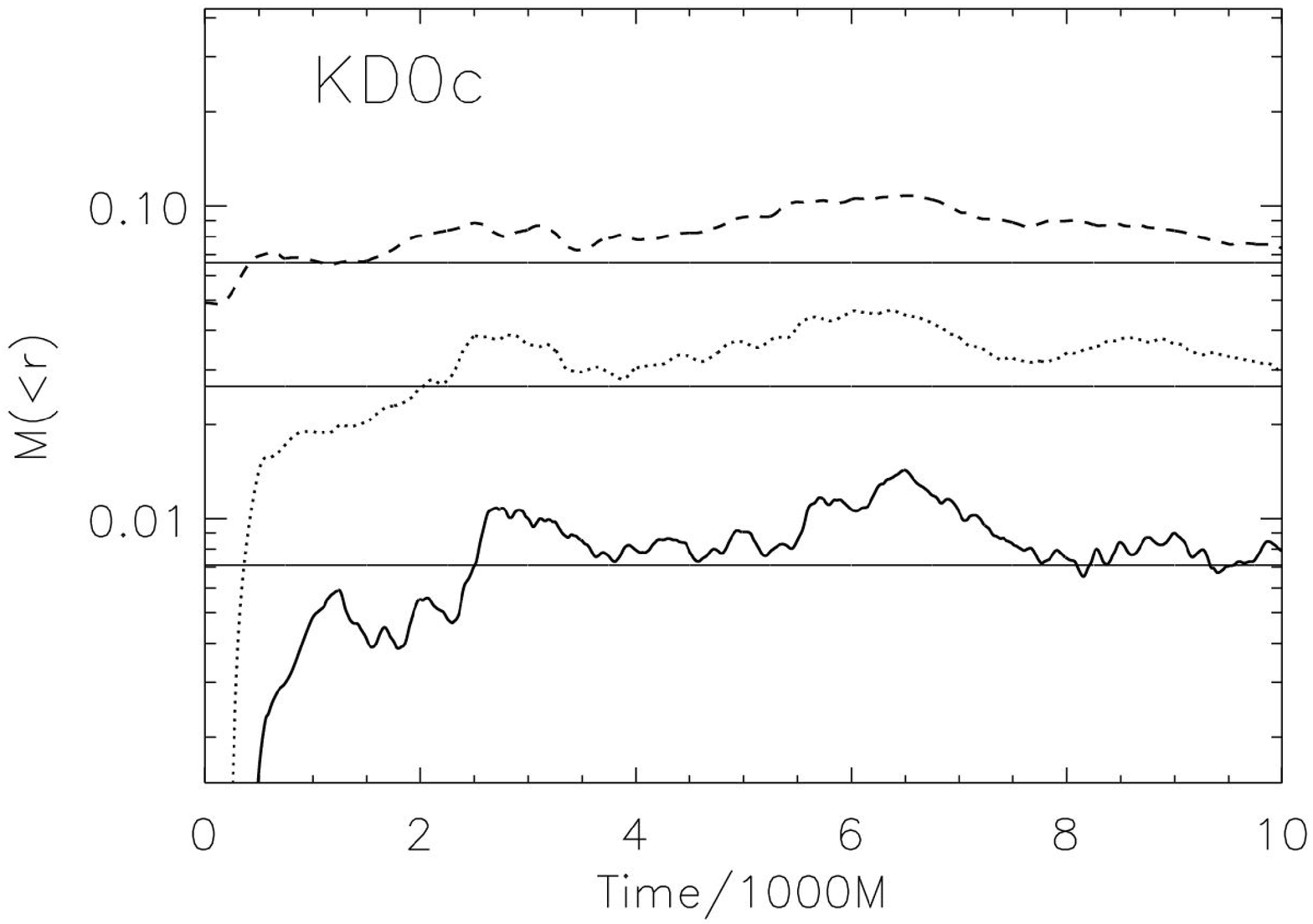}
\includegraphics[angle=90,scale=0.35]{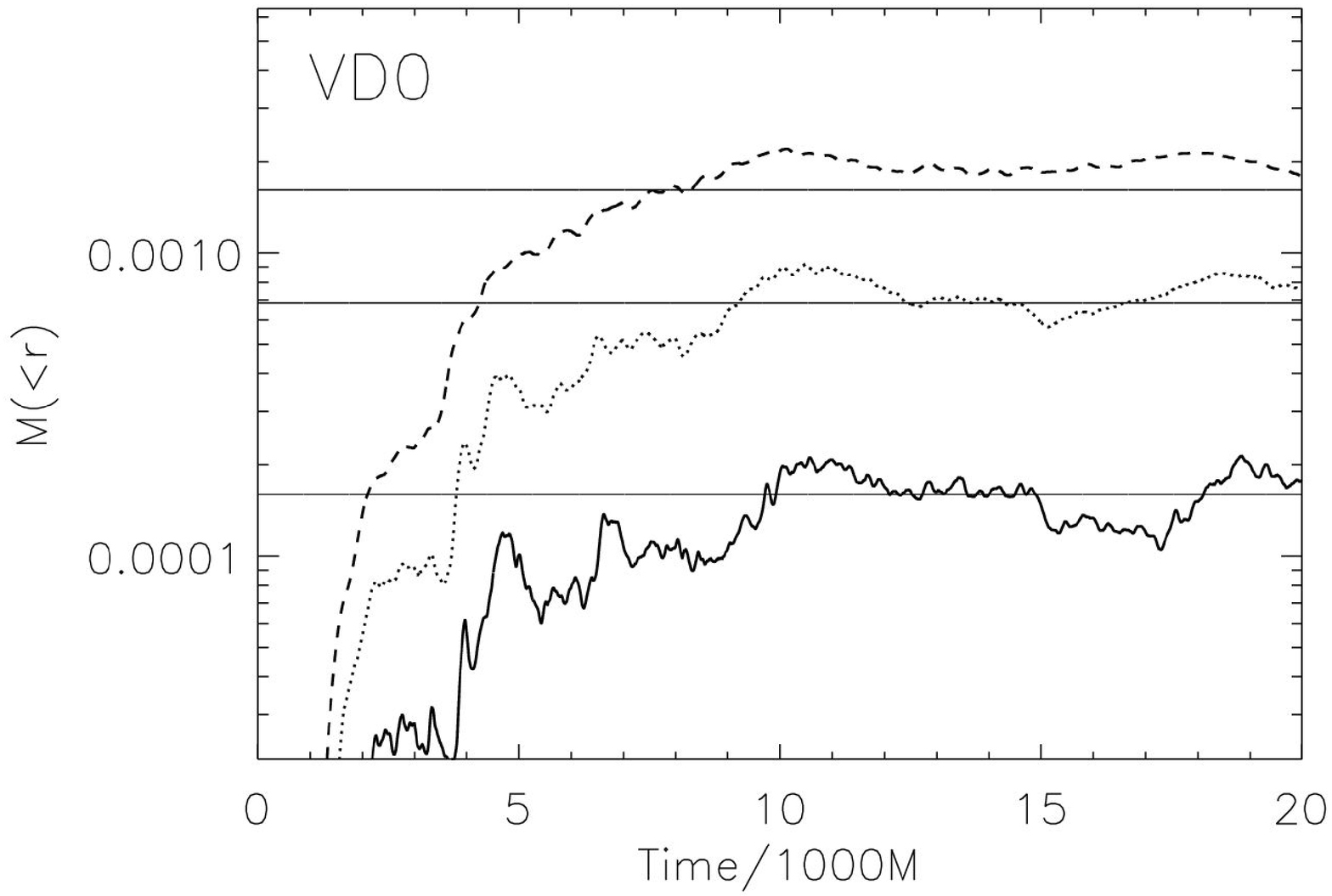}}
\caption
{Mass interior to $r =10M$ (thick solid curve), $r =15M$ (dotted curve),
and $r =20M$ (dashed curve) for the two GRMHD simulations.  The three
horizontal thin solid lines show $90\%$ of the final mass for each of
these radii.  (Left) KD0c.  (Right) VD0.
\label{fig:KD0VD0massfillin}}
\end{figure}

Similar data for the two GRMHD simulations is given in 
Figure~\ref{fig:KD0VD0massfillin}.  Both simulations reach inflow equilibrium
with respect to this criterion, after $\simeq 3000M$ in the case of KD0c,
after $\simeq 10000M$ in the case of VD0.

\begin{figure}
\centerline{\includegraphics[angle=90,scale=0.25]{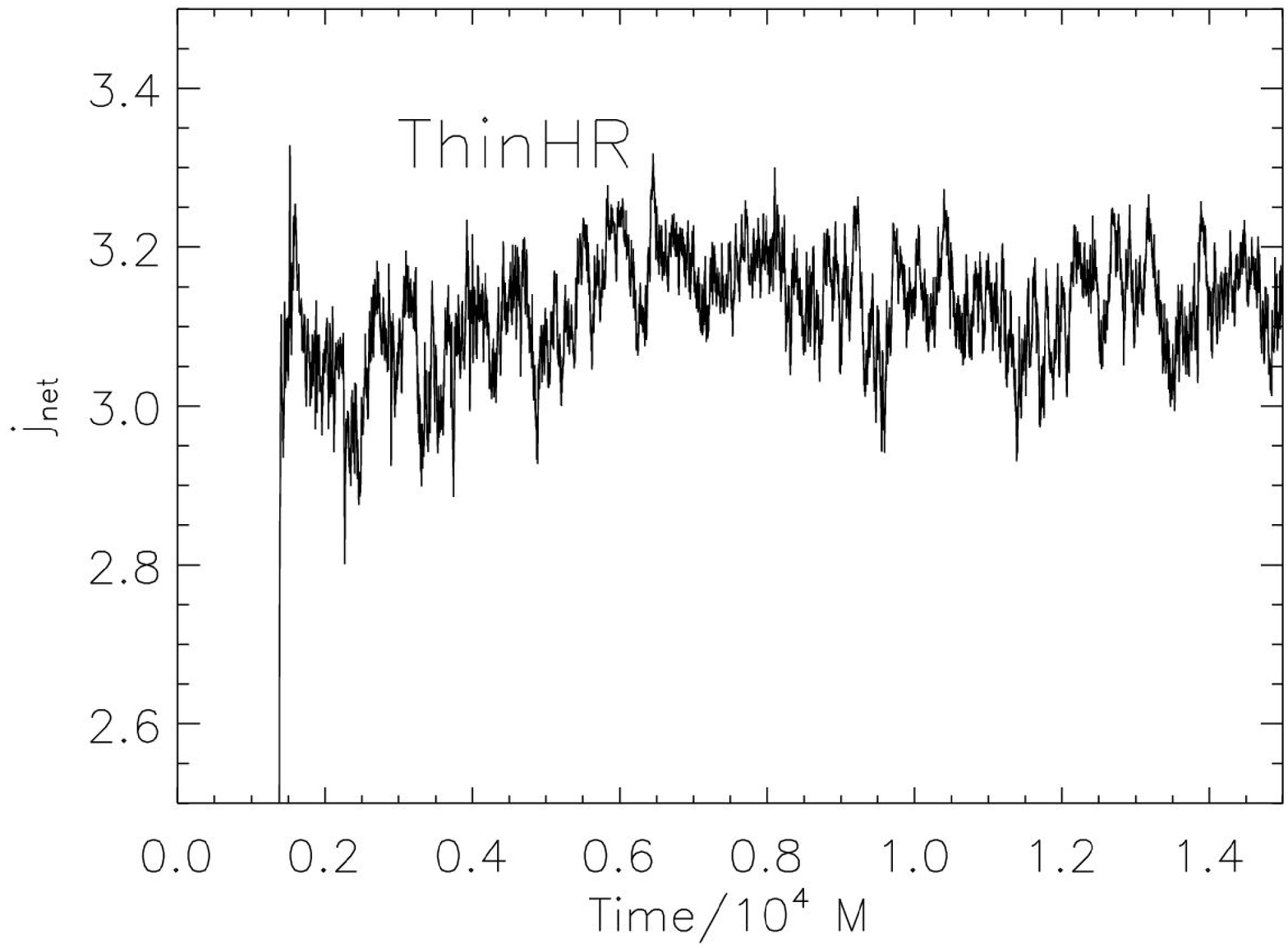}
\includegraphics[angle=90,scale=0.25]{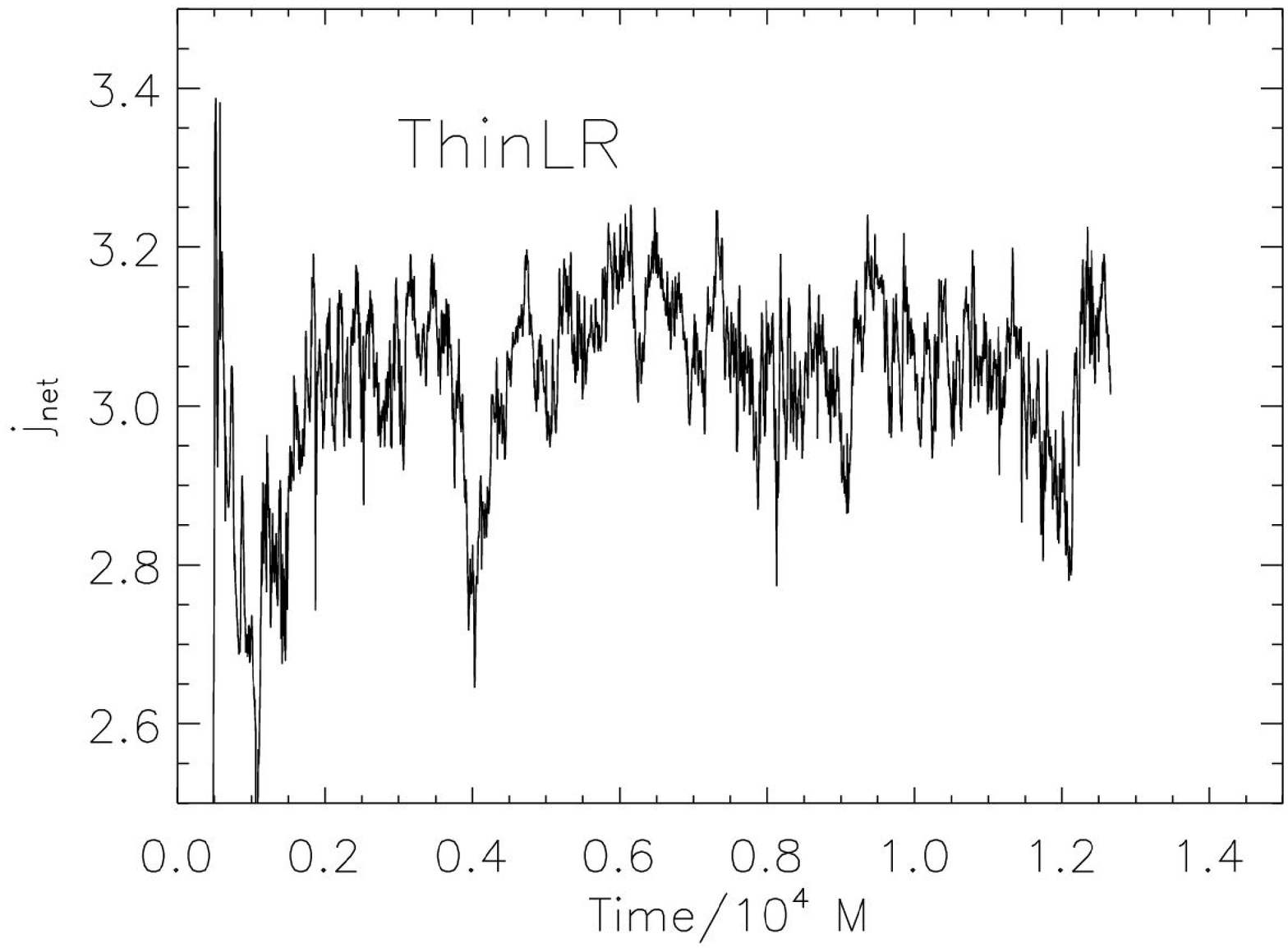}
}
\centerline{\includegraphics[angle=90,scale=0.25]{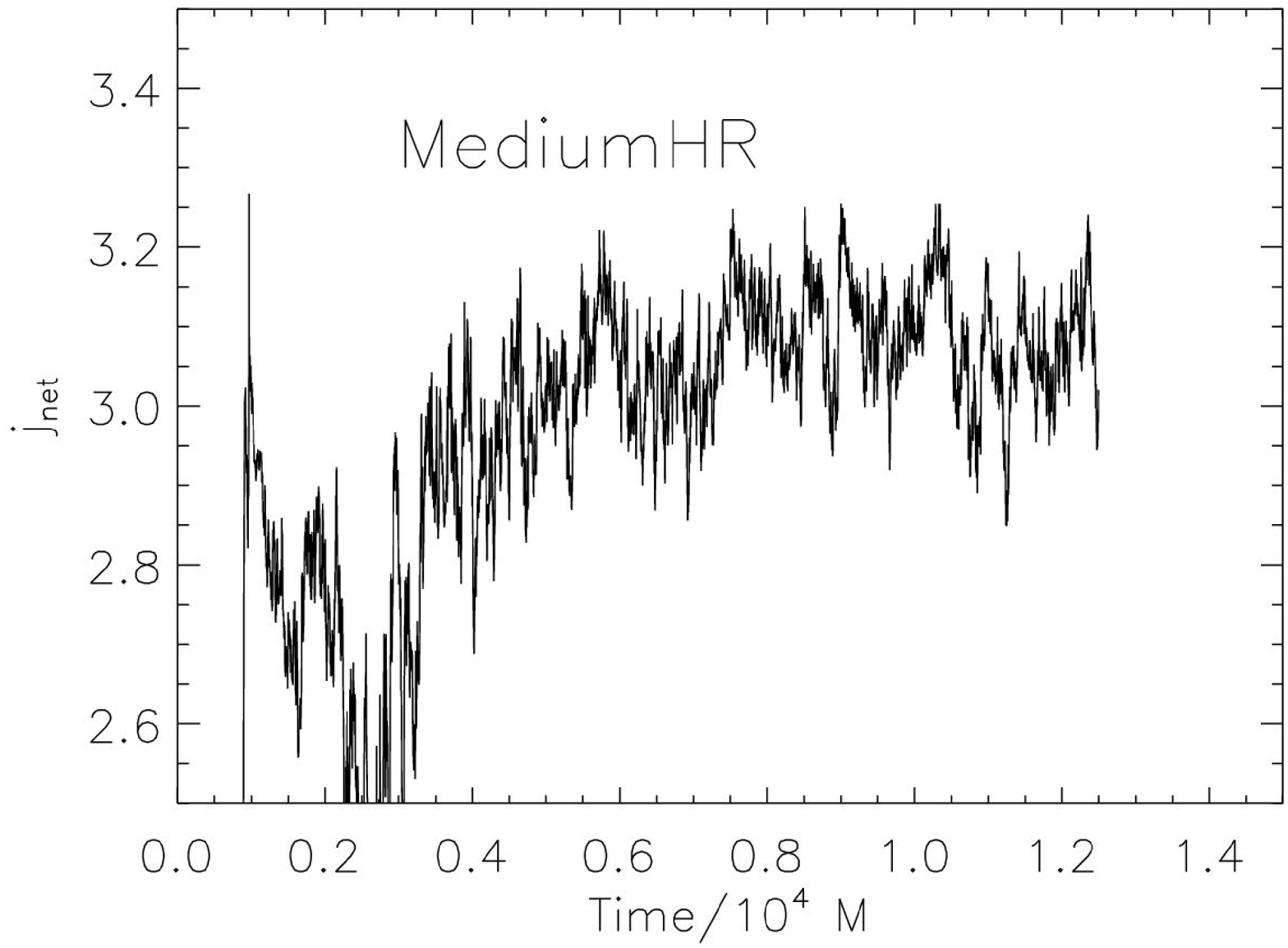}
\includegraphics[angle=90,scale=0.25]{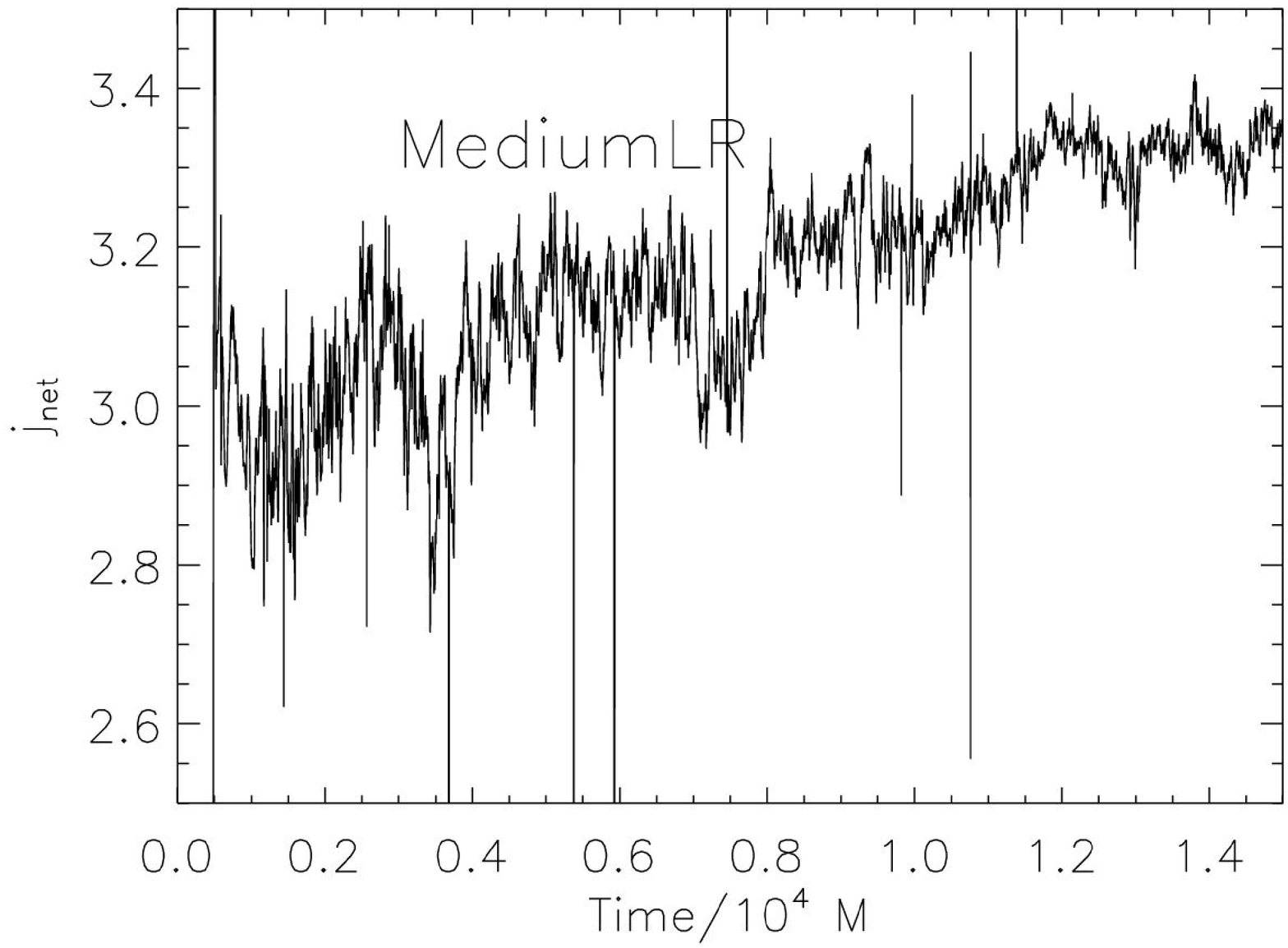}
\includegraphics[angle=90,scale=0.25]{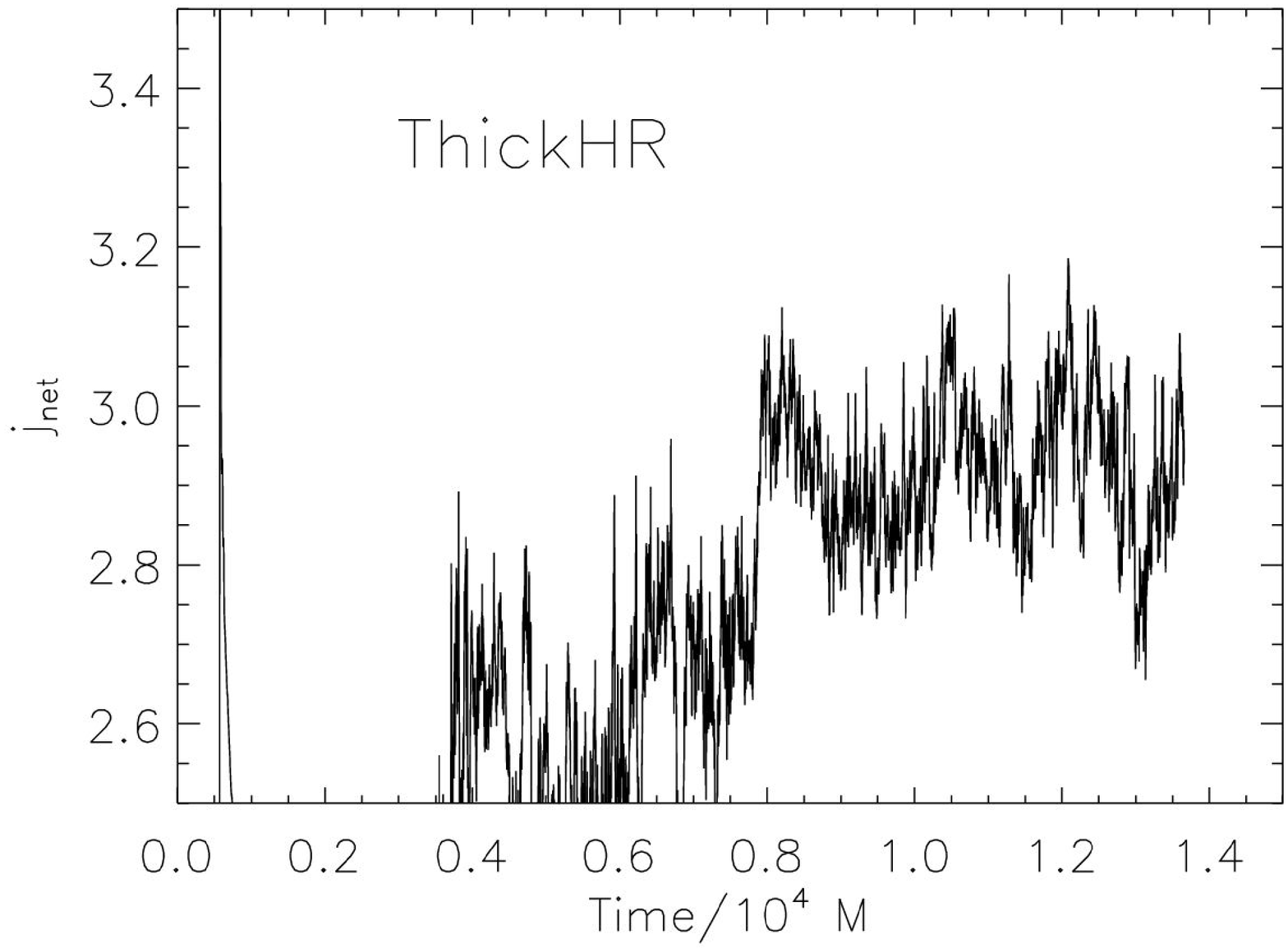}
}
\caption
{The net specific accreted angular momentum, $j_{\rm net}$, as a function
of time in ThinHR (top left), ThinLR (top right), MediumHR (bottom
left), MediumLR (bottom middle), and ThickHR (bottom right).  Note that
$u_\phi(\rm{ISCO}) = 3.464$ in Schwarzschild spacetime.
\label{fig:jnethist}}
\end{figure}

Another test of inflow equilibrium is provided by the history of the
specific angular momentum accreted into the black hole, $j_{\rm net}$.
This is determined by dividing the total angular momentum flux by the
accretion rate: 
\beq{
j_{\rm net}(r,t) \equiv \frac{ \langle {T^r}_\phi \rangle }{ \langle \rho u^r \rangle }
\quad , 
}
where the brackets represent the shell integration of the bracketed quantity:
\beq{ 
\langle X \rangle \equiv \int d\theta \, d\phi \, \sqrt{-g} \, X \quad . 
}
The quantity $j_{\rm net}(r,t)$ will be constant in $r$
where there is inflow equilibrium.  If the
accretion disk is truly in a statistically steady state in regard to
its angular momentum flow, this quantity should exhibit no trends in time,
varying only modestly due to fluctuations intrinsic to the turbulence.
Figure~\ref{fig:jnethist} shows $j_{\rm net}(\rhor)$ as a function of time for
the entire duration of each of the five new simulations.  They display
quite different behavior.  Once accretion begins, $j_{\rm net}$ hardly
varies in ThinHR.   Although there are no secular trends in ThinLR after
$t=2000M$, there are two sharp drops, at $t=4000M$ and $12000M$, and its
fluctuations are in general rather larger than in ThinHR.  The curve of
$j_{\rm net}(t)$ in MediumHR is more or less flat after $t=5000M$, but its
fluctuations are almost as large as those in ThinLR.  By contrast, $j_{\rm
net}$ rises in a succession of steps in MediumLR, showing relatively brief
periods of near-constancy.  Finally, $j_{\rm net}(t)$ rises rapidly from
the beginning of ThickHR until about $t=8000M$, holding approximately
steady from then until the end of the simulation at $13660M$.

The time-dependence of $j_{\rm net}$ in MediumLR is instructive.  Overall,
especially after
$10000M$, $j_{\rm net}$ rises toward $u_\phi(\rm{ISCO})$.  As discussed in
\S\ref{subsec:resolution}, the average number of grid zones per most
unstable wavelength dropped in this simulation as the field
weakened.  Inadequate resolution leads to ever-weaker magnetic field
which, in turn, results in a rise of the specific accreted angular
momentum toward the ISCO value as the magnetic stress is reduced.

\begin{figure}
\centerline{\includegraphics[angle=90,scale=0.25]{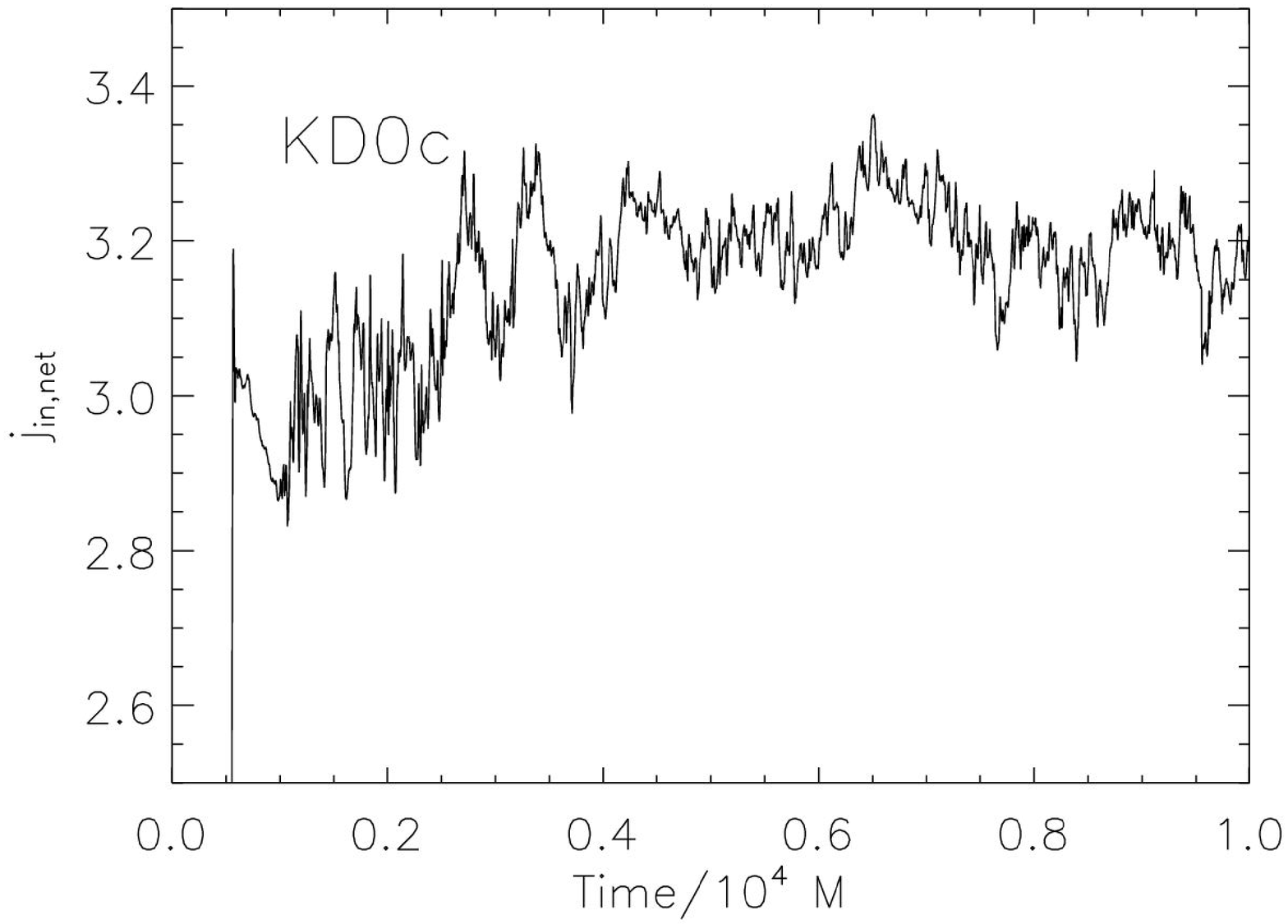}
\includegraphics[angle=90,scale=0.25]{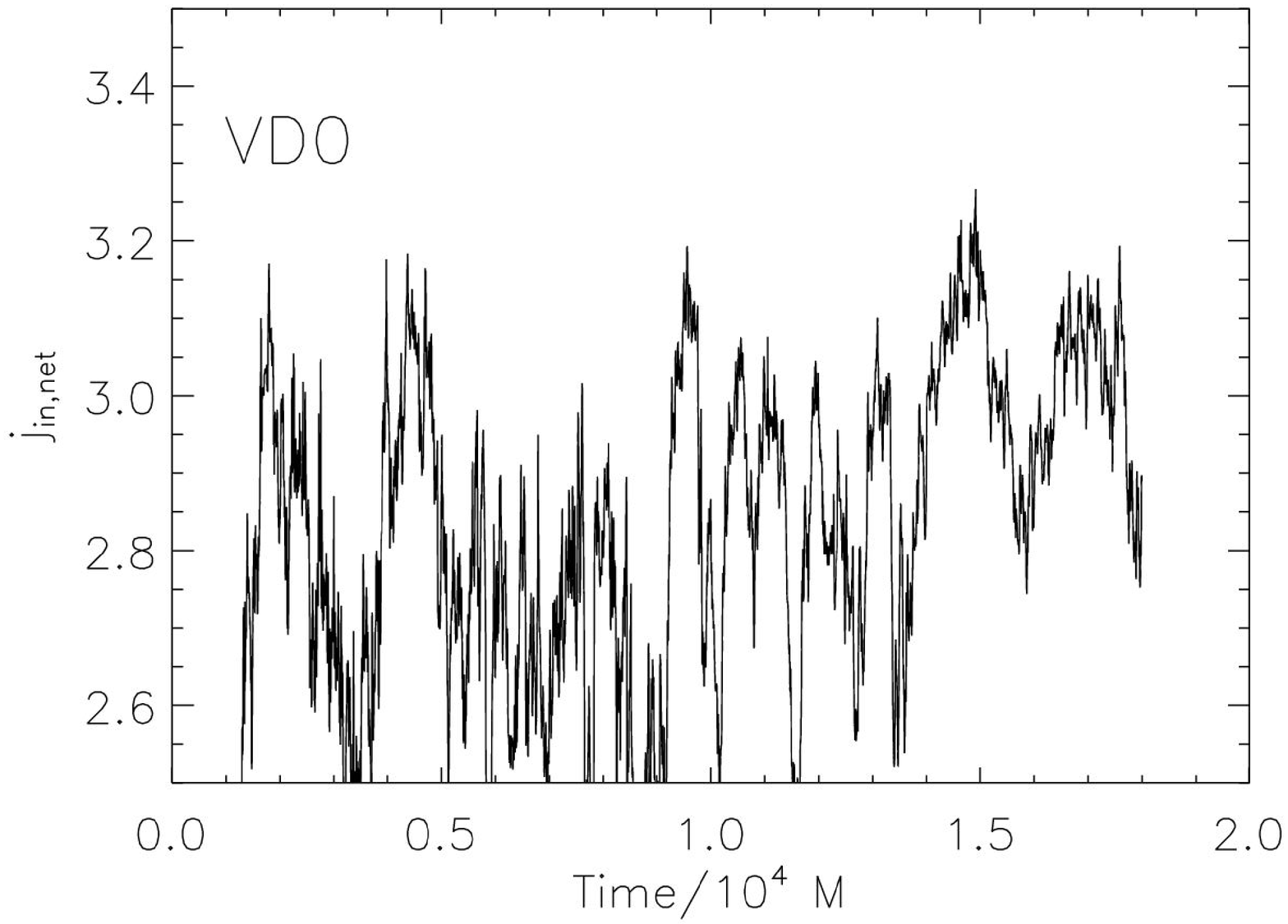}}
\caption
{The net specific accreted angular momentum, $j_{\rm net}$, as a function
of time in KD0c (left) and VD0 (right).  Note that $u_\phi(\rm{ISCO}) = 3.464$
in Schwarzschild spacetime.
\label{fig:KD0VD0jnethist}}
\end{figure}

Figure~\ref{fig:KD0VD0jnethist} shows $j_{\rm net}$ for the GRMHD
simulations KD0c and VD0.  The zero net-flux simulation, KD0,
shows no secular trend in this quantity after $4000M$, suggesting an
inflow equilibrium.  In VD0, on the other hand, the
fluctuations in $j_{\rm net}$ are much stronger,
and there also appears to be a rising trend from the beginning of the
simulation up until $t \simeq 14000M$, after which the trend flattens
out and the fluctuations begin to diminish.

Combining what we have seen in the mass-interior plots with those in
the $j_{\rm net}$ histories, we define the averaging periods,
$\Delta t_{\rm ave}$,  for these
simulations as the time when {\it both} criteria for inflow equilibrium
are met.  The results of this analysis are shown in the last column of 
Table~\ref{tab:simresults}.  In two cases (MediumHR, ThickHR),
the two tests single out the same periods; in one case (ThinHR), the
mass-interior equilibrium period is a portion of the specific angular
momentum equilibrium period; in four other cases (ThinLR, MediumLR, KD0c,
VD0), only a part of the period that meets the mass-interior test is
also in equilibrium according to the specific angular momentum test.

With the appropriate time-averaging period chosen, we can study how
the time-averaged accretion rate varies with radius.  Of the seven
simulations, in only one (ThinHR) is the contrast in $\langle \dot M\rangle=\langle \rho u^r \rangle$
inside $r =20M$ as much as $30\%$; in one (MediumLR) it is $\simeq 10\%$;
in all the others (ThinLR, MediumHR, ThickHR, KD0c, VD0), it is no more than
a few percent.

\subsection{Scale-height regulation}

\begin{figure}
\centerline{\includegraphics[scale=0.25]{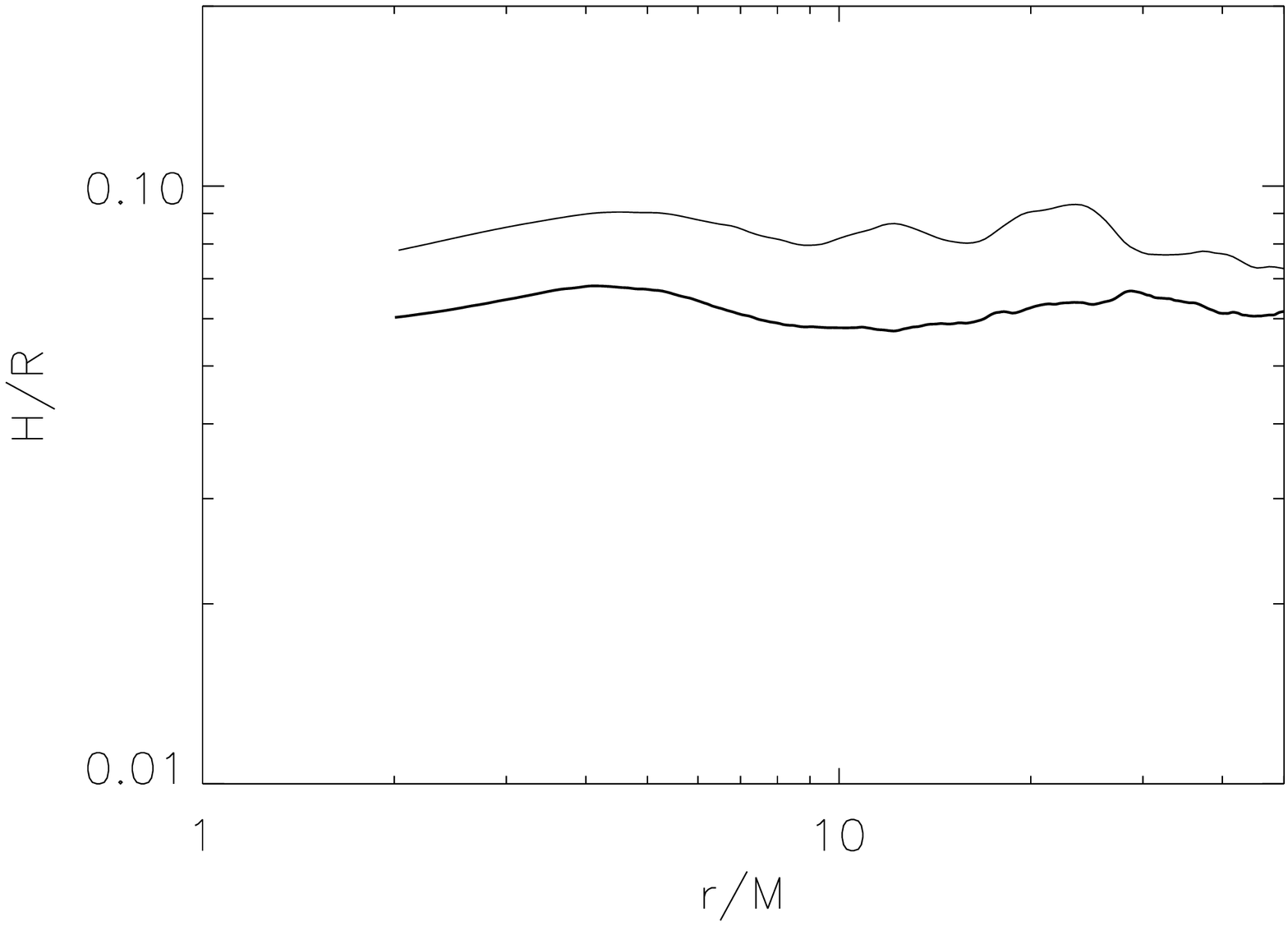}
\includegraphics[scale=0.25]{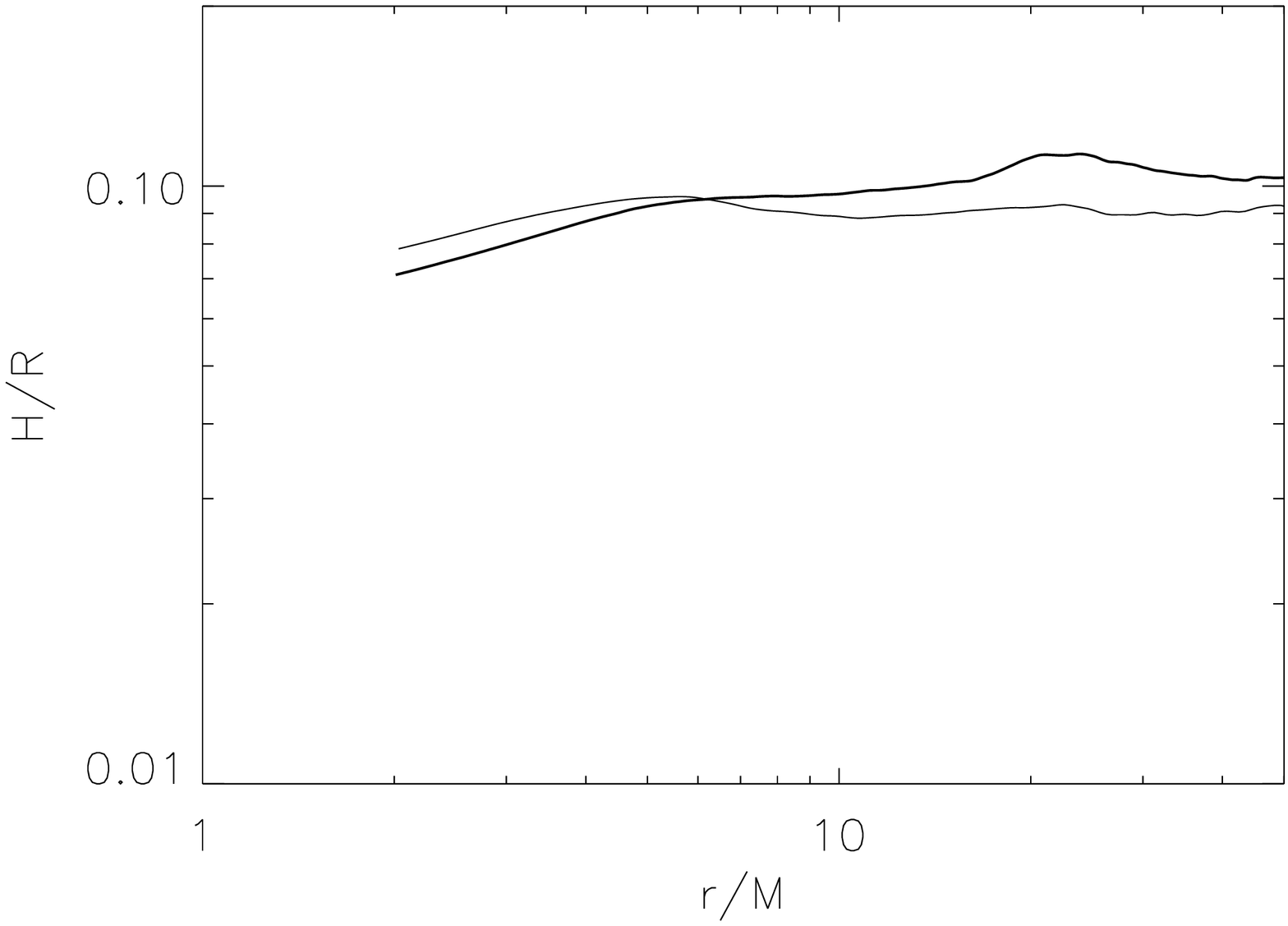}
\includegraphics[scale=0.25]{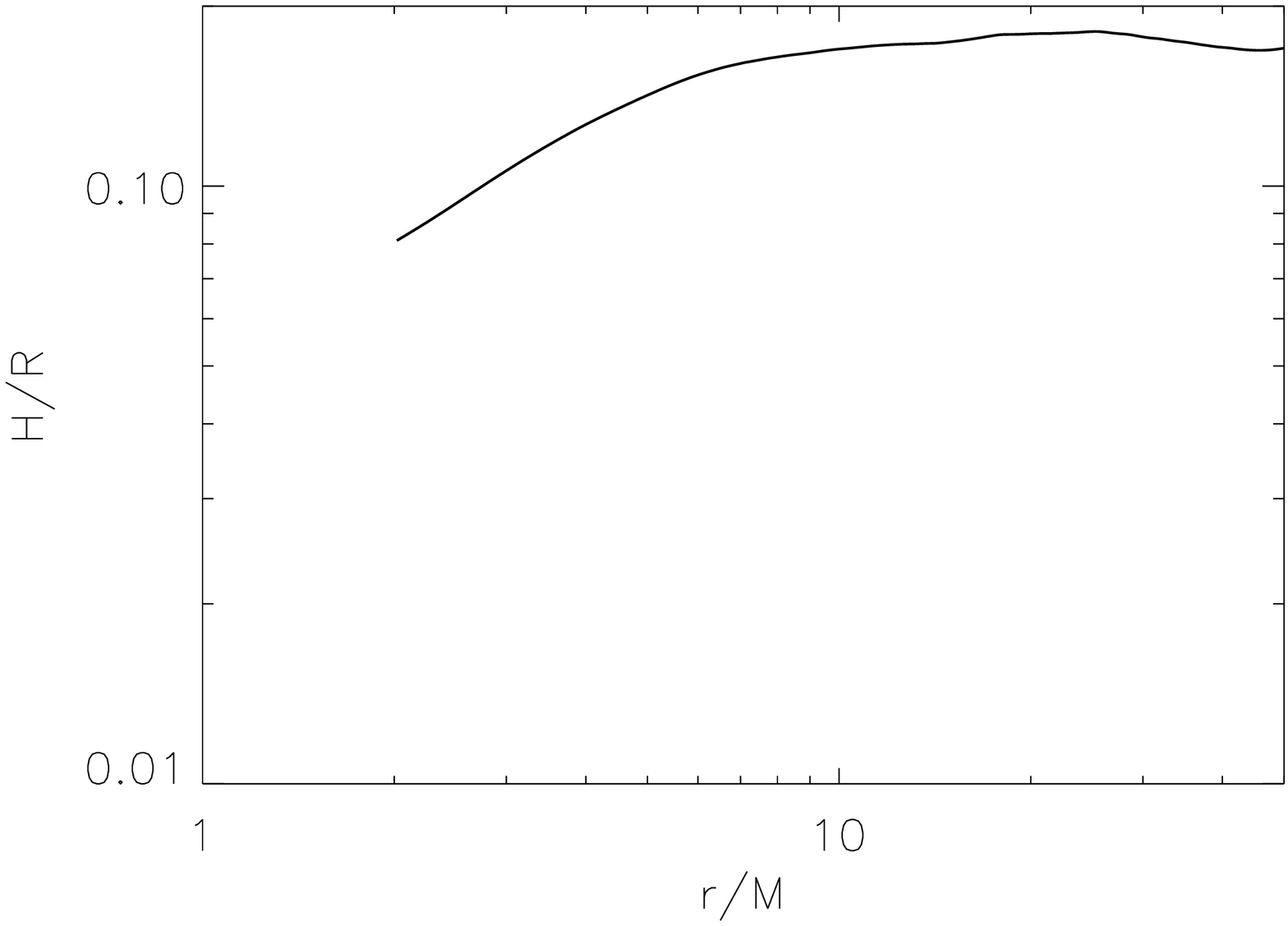}}
\caption
{Time-averaged scale height as a function of radial coordinate for each of
the HARM3D simulations. (Left) ThinHR and ThinLR,
(center) MediumHR and MediumLR, (right) ThickHR.  
In each case, the heavy curves correspond to HR,
the light curves to LR.
\label{fig:scaleheights}}
\end{figure}

Finally we examine the actual time-averaged scale heights achieved in the
various simulations; these are shown in Figure~\ref{fig:scaleheights}.
The scale-height regulation employed in the HARM3D simulations is quite
successful at enforcing a fixed ratio $H/R$ (except in the plunging
region in ThickHR), but, as shown in Table~\ref{tab:simresults}, the
actual value obtained can be different from the target, $\simeq 20\%$
greater in the cases of ThinHR, MediumHR, and MediumLR, but $\simeq 70\%$
greater in the case of ThinLR.  Part of this consistent offset can be
attributed to magnetic support, and part to the fact that the temperature
is typically slightly greater than the target.  The large offset in
ThinLR is a consequence of its initial condition, in which the gas was
given a thickness almost twice as great as the target.  Although its
initial mean plasma $\beta$ was 100, when cooling compresses the disk
by a sizable factor, the magnetic field strengthens while the gas
pressure falls.  As a result, much of the disk mass of ThinLR outside a
thin midplane layer was supported magnetically, and its scale height was
substantially increased beyond what its gas pressure alone could support.
This fact emphasizes the importance of choosing initial conditions relatively
close to the expected time-averaged state.  Inside the ISCO, the aspect
ratio decreases slightly because the inflow time is so short that the
disk cannot maintain vertical hydrostatic equilibrium.

\begin{figure}
\centerline{\includegraphics[scale=0.25]{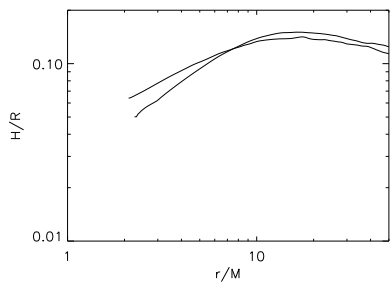}}
\caption{Time-averaged scale height as a function of radial coordinate for the
two GRMHD simulations.  The thick curve is the zero net-flux case, KD0c; the
thin curve is the non-zero flux case, VD0.
\label{fig:GRMHDscaleheights}}
\end{figure}

The time-averaged scale heights of the two GRMHD simulations are shown
in Figure~\ref{fig:GRMHDscaleheights}.  This code evolves the internal
energy equation, rather than the total energy equation,
and its simulations did not employ an explicit cooling
function.  Rather, entropy is conserved except where local shock heating
is captured by an artificial viscosity term.  Thus the scale height is not
controlled directly and is determined primarily by the initial condition.
In the end, the two simulations have very similar scale-height profiles,
with $H/R \simeq 0.15$ for $r \geq 10M$, but declining inward, reaching
$\simeq 0.12$ at the ISCO ($r =6M$) and $\simeq 0.06$ just outside the
horizon.  The mean $H/R$ for both is $\simeq 0.14$.

\section{Results: Stress in the Inner Disk}

With this background in mind, we can now discuss the results for
time-averaged stress in the inner disk.  We will present them in two
ways: in terms of the radial profile of the spherical shell-integrated
fluid-frame electromagnetic stress, and in terms of the angular
momentum flux and the quantity $j_{\rm net}$ defined in the introduction.

\subsection{Fluid-frame electromagnetic stress profile}

We begin with the radial profile of the electromagnetic stress.  Both for the
purpose of highlighting the physics and for the purpose of contrasting with
the Novikov-Thorne model, it is best to compute it in the fluid frame:
\begin{equation}
W^{\left( r\right)}_{\left(\phi \right)}(r)  =  \frac{ \int \int dx^{\left(\phi\right)} dx^{\left(\theta\right)}
\left(|b|^2 u^\nu u_\mu - b^\nu b_\mu\right)
e^{\left( r\right)}_\nu e^\mu_{\left(\phi\right)}/(4\pi)}
{\int dx^{\left(\phi\right)} |_{\theta=\pi/2}}   \quad , 
\label{eq:flframestress}
\end{equation}
where $b^\mu$ is the magnetic 4-vector.  Each component of the
vector $dx^{\left(\mu\right)} = {e^{\left(\mu\right)}}_{\nu} \,
dx^\nu$ represents the extent of a cell's dimension as measured in
the fluid element's rest frame, and ${e^{\left(\mu\right)}}_{\nu}$ is
the orthonormal tetrad that transforms vectors in the Boyer-Lindquist
coordinate frame to the local fluid frame (see \cite{BHKedge}
for explicit expressions for the tetrad).  The vector $dx^\nu$ is the
Boyer-Lindquist coordinate frame version of the Kerr-Schild vector
$dx^\nu_\mathrm{KS} = 
\left[0,\Delta r, \Delta \theta, \Delta \phi\right](r,\theta,\phi)$, 
where $\Delta r$, $\Delta \theta$, $\Delta \phi$ are the radial,
poloidal and azimuthal extents of our simulation's finite volume cell
located at $(r,\theta,\phi)$\footnote{Care must be exercised to properly
evaluate volume integrals in the fluid frame.  Both \cite{KHH05} and
\cite{Shafee08} correctly projected the stress tensor into the fluid
frame, but failed to similarly project the volume element.}.

The physical significance of the electromagnetic fluid-frame stress profile is that
it describes the rate at which angular momentum is carried outward by
electromagnetic fields.  The net angular momentum accreted by the black
hole is diminished to the degree that stresses like these convey angular
momentum outward even while inflowing matter carries its orbital angular
momentum inward.   The Novikov-Thorne model assumes that the stress begins
to decline outside of the ISCO, reaching zero at that point.  \cite{ak00}
showed how changing that boundary condition to account for non-zero stress
at the ISCO, i.e., $j_{\rm net} < u_\phi(\rm{ISCO})$, could significantly
alter the shape of the stress profile.
Even in the disk body, well outside the ISCO, a smaller $j_{\rm net}$
can lift the time-averaged stress above the Novikov-Thorne curve.

\begin{figure}
\centerline{\includegraphics[angle=90,scale=0.4]{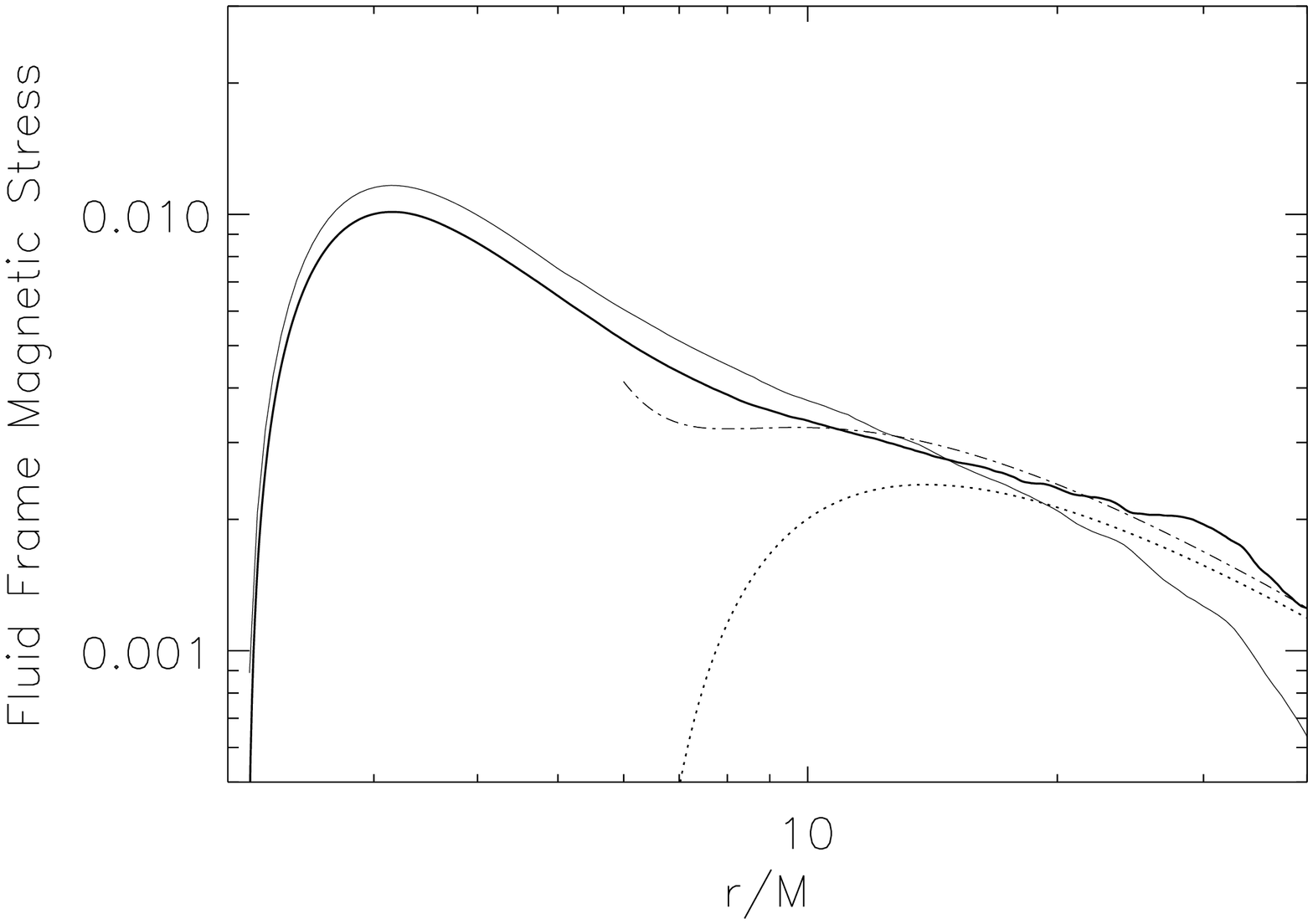}
\includegraphics[angle=90,scale=0.4]{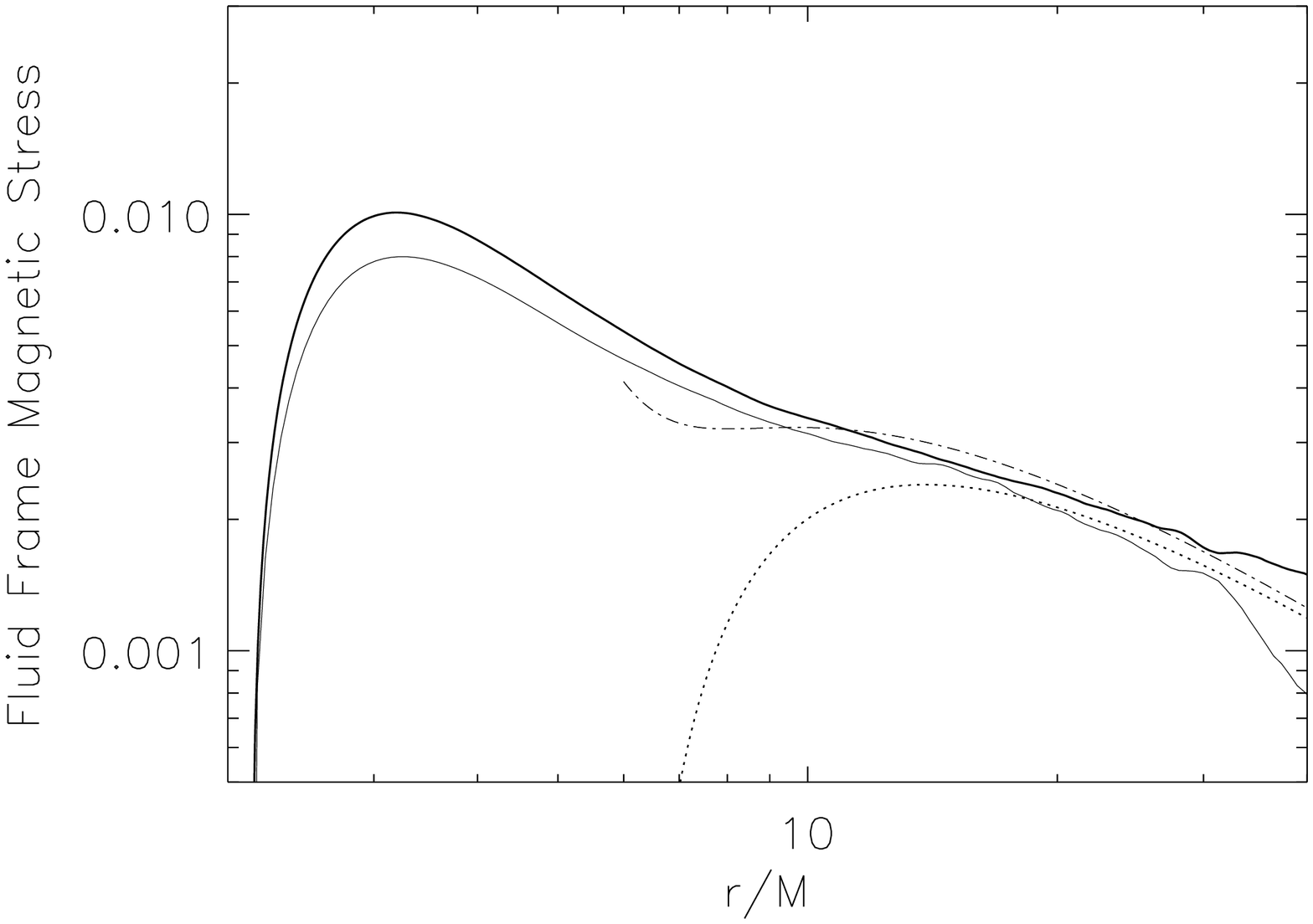}}
\centerline{\includegraphics[angle=90,scale=0.4]{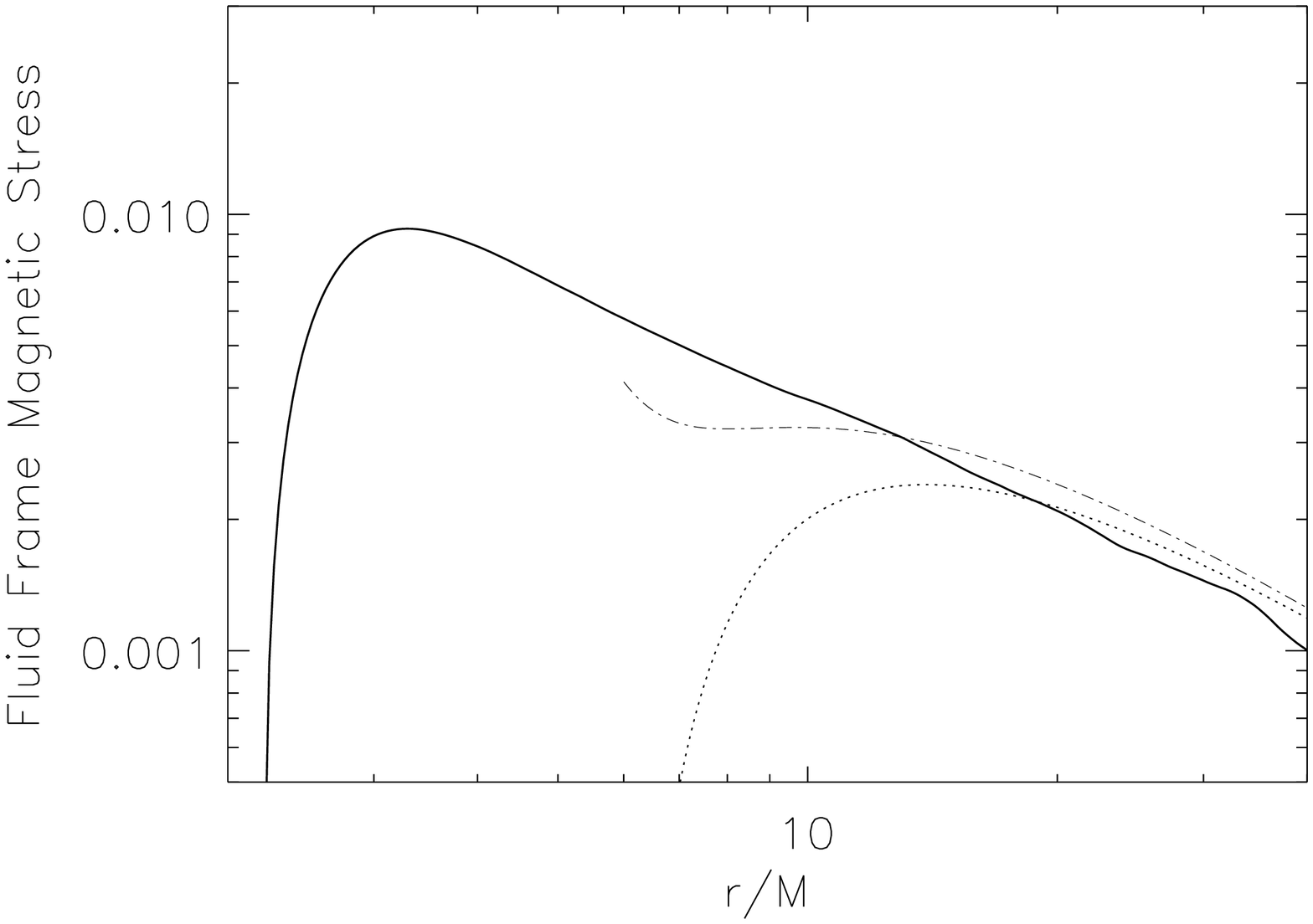}}
\caption{
Fluid-frame electromagnetic stress, normalized by that simulation's mean
accretion rate, for each of the disk aspect ratios.
(Top left) ThinHR and ThinLR, (top right) MediumHR and MediumLR, (bottom) ThickHR.  
Heavy and light curves correspond respectively to HR and LR.  Dotted curves show the
Novikov-Thorne model's prediction, dot-dashed curves the
prediction of the Agol-Krolik model with additional efficiency 
$\Delta\epsilon = 0.015$---a $26\%$ increase relative to the Novikov-Thorne 
efficiency $\epsilon = 0.057$.
\label{fig:stressprofile}}
\end{figure}

Figure~\ref{fig:stressprofile} shows the time-averaged
$W^{\left(r\right)}_{\left(\phi \right)}(r)$ for each of the HARM3D simulations,
normalized by their time-averaged accretion rate.  The dotted line
shows the Novikov-Thorne prediction and the dot-dash curves are
examples of an Agol-Krolik profile \citep{ak00}.  
Remarkably, all five of the HARM3D
simulations show almost identical profiles, differing only in very minor ways.
We concentrate on the region where inflow equilibrium applies, here
$r \leq 20M$.  Outside the ISCO, but inside the domain of inflow
equilibrium, the fluid frame stress usually (but not always) lies slightly
above the Novikov-Thorne prediction.  Although the match is not perfect,
it is somewhat better described by the Agol-Krolik model.  As the flow
approaches the ISCO, where the Novikov-Thorne model would predict that
the stress begins to fall, the electromagnetic stress rises steadily.
Inside the ISCO, in the plunging region, the stress continues to rise
inward, with a slope that is similar to or slightly steeper than outside
the ISCO.  Just outside the event horizon, the stress falls sharply to
zero: because this stress component is nothing more than the radial flux
of angular momentum of rotation in the disk plane, when the black hole
has no angular momentum (i.e., does not rotate), it cannot act as a source
of angular momentum and the stress immediately outside it must go to zero.

The corresponding profiles for the two GRMHD simulations are shown in
Figure~13.  They are very similar to one another, and qualitatively similar to,
but quantitatively different from, the HARM3D profiles.  Like the HARM3D profiles,
the radial slope of the stress is nearly constant in the disk outside the
ISCO; unlike the HARM3D profiles, in these two the stress rises somewhat
{\it more} steeply inside the ISCO.  As a result, the peak stress in these
two simulations is $\simeq 2$ times greater than seen in the HARM3D cases.

In conclusion, despite the range of temperatures and scale heights
covered by these models, there is a great similarity between the stress
profiles of all the HARM3D models as well as KD0c and VD0.  In other words,
with respect to this measure of the stress, {\it there appears to be no
dependence on $H/R$ whatsoever.}

Furthermore, the similarity between KD0c and VD0 suggests that the
change of magnetic topology from closed dipolar loops on the scale of
the disk thickness to net vertical field also makes little difference to
the radial variation of accretion stress.  One possible explanation for
this insensitivity is the fact that reconnection events in the corona
largely decoupled the magnetic field in the inner disk of VD0 from the
large-scale flux \citep{BHK09}.  The remaining field in the disk then has a topology
not so different from that in the other simulations.  In
\S\ref{sec:pastresults} we review the results from previously
published simulations to explore further the possible role of field
topology on ISCO stress.

\subsection{Specific accreted angular momentum}

The value of $j_{\rm net}$, the mean angular momentum accreted per
unit rest-mass, summarizes the net angular momentum flow in the system.
It is determined by several effects.  In the accretion disk body, the
orbital angular momentum, $u_\phi$, is close to the value associated with
a circular test-particle orbit at that radius, but can be altered by an
amount $\sim (H/R)^2$ by radial pressure gradients, both gas and magnetic.
Stresses, both electromagnetic (Maxwell) and fluid (Reynolds) move
angular momentum through the accretion flow; to the degree that they
have a net divergence, they can either add or remove angular momentum
from the fluid.  In the classical Novikov-Thorne model,
the fluid's angular momentum is {\it assumed} to match
the local circular orbit angular momentum at all locations outside the
ISCO, but is fixed at the ISCO angular momentum at all smaller radii.
The stresses, whether Maxwell or Reynolds, are constrained to be exactly
what they need to be to produce this result: finite in the disk body,
zero in the plunging region.  As a result, $j_{\rm net}$ is predicted
to be exactly $u_\phi (\rm{ISCO})$.

As can be seen from the data listed in Table~\ref{tab:simresults}, this
is not the case in the simulations.  We find that, over the range of
thicknesses and magnetic geometries studied, $j_{\rm net}$ ranges from
$\simeq 2.93$ to $\simeq 3.21$.  Thus, in all these cases, the accreted
angular momentum per unit accreted rest-mass is 7--$15\%$ below 
$u_\phi(\rm{ISCO})=3.464$.  Interestingly, the largest $j_{\rm net}$ by far was
seen in KD0c, a simulation performed on a comparatively coarse grid;
if it were discounted, the depression of $j_{\rm net}$ below $u_\phi$
would be 10--$15\%$.

The separate elements contributing to this departure from
the Novikov-Thorne prediction are shown by the curves in
Figures~\ref{fig:netangmom} and \ref{fig:KD0VD0netangmom}.  The solid
lines correspond to the net specific angular momentum flux; in inflow
equilibrium this should
be constant with radius.  The dashed curves show the
time-averaged specific angular momentum carried by the accreting matter,
i.e., $\langle \rho h u^r u_\phi\rangle/\langle \rho u^r \rangle$.
The difference between these curves and the net angular momentum flux is due
to the electromagnetic stress.  The dot-dash curves show the time-averaged
mass-weighted mean angular momentum at each radius, $\langle \rho h
u_\phi \rangle/\langle \rho \rangle$.  These last two quantities can
be compared with
the $u_\phi$ corresponding to a circular orbit, shown by the dotted line
(which is held constant at the ISCO value inside of that radius).

For all the models, the mass-weighted mean angular momentum generally
follows the circular orbit value outside of the ISCO, but continues to
decline inside of that radius rather than holding steady.  In the thin
disk models the offset from the circular orbit value is small, while in
the thicker cases this offset is somewhat larger.  This is precisely
what one would expect; the gas in the disk is partially supported by
the outward radial decline in pressure, primarily magnetic, in disks
with a larger $H/R$.

In every case, the curve of mean angular momentum accreted by the matter
lies well below the curve of the local mass-weighted mean angular
momentum.  One way to view the origin of this offset stems from the
fact that the material in the disk is turbulent;
local fluid elements have angular momenta that fluctuate from time
to time and from place to place.  It should be no surprise that fluid
elements with angular momentum slightly smaller than the value that
would support a local circular orbit tend to move inward faster than
those with larger angular momentum.  In other words, the mean accreted
angular momentum is systematically biased toward lower values by orbital
mechanics that sorts the fluid elements according to their place in the
local angular momentum distribution.

This effect can also be identified with turbulent Reynolds stress.  To see this
identification directly, consider the equation of angular momentum equilibrium
integrated over spherical shells and averaged in time:
\begin{equation}\label{eq:totangmomcons}
\langle \rho h u^r u_\phi \rangle + \langle M^r_\phi \rangle 
         = j_{\rm net} \langle \rho u^r \rangle,
\end{equation}
where $M^r_\phi$ is the Maxwell stress.  The term $\propto u^r u_\phi$ can be broken
into two pieces, one reflecting the advection of the mean angular momentum (weighted
by enthalpy), the other reflecting departures from the mean.  Dividing
through by the mass accretion rate, the previous equation becomes
\begin{equation}\label{eq:normangmomcons}
\langle h u_\phi \rangle - \frac{\langle \delta (\rho u^r)\delta (h u_\phi) \rangle
         + \langle M^r_\phi \rangle}{-\langle \rho u^r \rangle} = j_{\rm net},
\end{equation}
where $\delta X \equiv X - \langle X \rangle $.  That is, the net
rate at which angular momentum is carried inward per unit rest-mass
accreted is the local mean angular momentum reduced by the ratio of
the total stress, Maxwell plus turbulent Reynolds, to accretion rate.
Comparing this formalism to Figures~\ref{fig:netangmom} and
\ref{fig:KD0VD0netangmom}, we see that the dot-dashed curves show 
$\langle \rho h u_\phi \rangle / \langle \rho \rangle \simeq \langle h u_\phi \rangle$, 
while the dashed curves show 
$\langle \rho h u^r u_\phi\rangle/(-\langle \rho u^r\rangle)$.
Their offset can then be
attributed to the turbulent Reynolds stress normalized to the accretion
rate, $\langle (\delta \rho u^r)\delta (h u_\phi) \rangle/(-\langle \rho
u^r\rangle)$.  This turbulent Reynolds stress can be quantitatively
significant, particularly in VD0 and to a lesser degree in ThickHR.

The separation between the curve of the accretion-weighted mean angular
momentum and $j_{\rm net}$ is the electromagnetic angular momentum
flux, and in every simulation but VD0 it clearly makes the largest
contribution to the outward angular momentum flux.  In all the other cases,
the only place where the Maxwell stress does not outweigh the Reynolds
stress is in the immediate vicinity of the event horizon.  There, the
matter's angular momentum flux becomes almost exactly the total because
a non-spinning black hole has no angular momentum to lose.  For the same
reason, just outside the horizon $||b||^2 u^r u_\phi$ comes to exceed
$b^r b_\phi$ in magnitude, and the net electromagnetic angular momentum
flux turns (weakly) negative.

To conclude, then, all these models show values of $j_{\rm net}$ that are
reduced below the ISCO value, regardless of $H/R$.  The electromagnetic stress
in the fluid frame hardly varies at all from one simulation to the next;
consequently, its contribution to the net angular momentum flux is likewise
nearly the same in all cases.  Scale height does seem
to have an effect on the run of $u_\phi$ through the disk: $u_\phi$ is
reduced below the circular orbit value in proportion to the magnitude
of radial pressure support.
It also appears that the Reynolds stress levels may be partially
controlled by the scale height in the disk and partially by the
magnetic topology.

\section{Review of Previous Simulations}
\label{sec:pastresults}

In this section we summarize some results from previous simulations.
Some of these, although not designed specifically to study the influence
of scale height, can nevertheless provide additional information about how
the inner disk stress depends on other parameters, notably magnetic
field topology.  Others are more directly comparable to our HR
series of simulations.

We have already discussed models KD0c \citep{KHH05} which began with
simple dipole loops in the initial torus, and VD0 \citep{BHK09},
a model that began with a vertical field piercing the initial torus.
\cite{BHK08} feature three simulations all computed in a Kerr ($a/M=0.9$)
spacetime with identical fluid initial states, but differing initial
magnetic fields: in KDPg, the field configuration was nested dipolar
loops similar to those in KD0c; in QDP, there was a pair of quadrupolar
field loops above and below the equator whose associated currents had
opposite signs; and in TDPa, the initial condition held only toroidal
field.

Figure~2 of \cite{BHK08} provides data for investigating how these
different initial field configurations affect inner disk stress.
Panel~f shows the coordinate-frame electromagnetic angular momentum flux
(shell- and time-averaged $T^r_\phi(EM)$) as a function of radius.
The curves for KDPg and QDPa, the large dipolar field and quadrupolar
simulations, are nearly identical.  Both curves are generally a factor of
several higher than that of TDPa, the one whose initial field was purely toroidal.
Figure~2c of \cite{BHK09} shows the radial run
of net specific angular momentum (what we call $j_{\rm net}$ here, but is
labeled $L$ in that figure).  Both KDPg and QDPa are slightly below
(by $\simeq 0.05$) the ISCO value of $2.1$.  TDPa, on the other hand,
has an average accreted specific angular momentum of 2.13, very nearly
the exact ISCO angular momentum, when averaged over
the last $9000 M$ in time.  The similarity between
quadrupole and dipole initial fields carries over to simulations with
a Schwarzschild hole.  Another GRMHD simulation, QD0 (described in
\citep{BHKedge}),
also has quadrupolar loops in the initial torus.  It was run for $10^4 M$
in time, and averaging over the last $2000 M$ in time gives a value of
$j_{\rm net} = 3.21$ for the specific angular momentum accreted into the hole,
similar to KD0c.  As a whole, these results suggest that the crucial
distinction may be between a field with significant poloidal character
(either with or without net flux) and one that is only toroidal.

There is another case where we can clearly see the significance of
a particular field configuration.  The vertical field model VD0 was
evolved in both two- and three-dimensions \citep{BHK09}.
In the present context the contrast between the two is interesting
\citep[see Fig.~7 of][]{BHK09}.  In the axisymmetric simulation
the value of $j_{\rm net}$ shows strong fluctuations.  Between the times of
$10^4 M$ and $1.5\times 10^4 M$ (the end of this simulation),
the mean of $j_{\rm net}$ is 2.85, but with a standard deviation of 0.43.
Values as low as $\sim 1.0$ are reached at particular moments.  The mean
value of $j_{\rm net}$ over the same interval in VD0 is similar, 2.89, but
the standard deviation is only 0.18 and the minimum value reached is 2.43.
Axisymmetric simulations with vertical fields typically show strong MRI
``channel modes'' characterized by extended radial flows accompanied
by radial magnetic field. It is the presence of those extended radial
fields through the plunging region that provide the strong torques.
Thus, the two-dimensional simulation illustrates the basic principle
obtaining in three-dimensions, but in exaggerated form.
In this particular case, the presence of a net vertical field (which
cannot be reconnected away within the disk) also prevents
the antidynamo effect from dissipating the turbulence, allowing the
stress in the plunging region to remain over the full evolution.

Our HARM3D simulations are very close, both in numerical technique and
parameters, to the general
relativistic MHD simulation of \citet{Shafee08}.  The Shafee et~al. simulation
was done, like ours, in the context of a non-rotating black hole.
Both employed intrinsically conservative Godunov algorithms differing
in only minor respects.  Shafee et~al. used a temperature regulation scheme
based on maintaining constant entropy rather than a target temperature, but---compared 
to our ThinHR simulation---the resulting aspect ratio was only slightly
thicker in the mean and somewhat less constant as a function of radius.

In their initial hydrodynamic conditions, our HR simulations were
almost identical to those of Shafee et~al., differing only in the
$q$ parameter (theirs was chosen to give an initial state with $H/R
\simeq 0.1$, ours had variously $H/R \simeq 0.05$, 0.08, and 0.16).
However, they did differ in the initial state of the magnetic field.
Whereas our initial magnetic field was a set of nested dipole loops centered
on the pressure maximum, they imposed two sets of loops, centered on $r=28M$
and $r=38M$, which they then perturbed randomly with $\simeq 50\%$
fractional amplitude.

The spatial grid used by Shafee et~al. was also very similar to the one we
employed for ThinHR.  They used a grid with $512 \times 128 \times 32 $ cells,
whereas ThinHR used $912 \times 160 \times 64$ cells.  Both radial grids
were logarithmic, but their grid extended to a slightly smaller radius
than ours ($50M$ as opposed to $70M$); their radial cell size was therefore
about 1.6 times larger than ours.  The polar-angle grid scheme
for ThinHR was finer than that of Shafee et~al. near the equator, resulting in
$\simeq30\%$ more cells per scale height.
The two simulations had identical azimuthal resolutions because, even
though we used twice as many cells, our azimuthal extent was also twice as
great ($\pi/2$ as opposed to $\pi/4$).
Shafee et~al. reported that in their initial condition, there were $\sim
10$ cells per fastest-growing MRI wavelength, but do not say how that
number changed during the simulation.

Despite this very close similarity, Shafee et~al.  arrived at a result
that was quite different from ours.  They found that the fluid-frame
electromagnetic stress followed the Novikov-Thorne prediction very
closely all the way to $r =9M$, and then maintained more or less that
amplitude all the way to $r \simeq 2.5M$.  In contrast, as
Figure~\ref{fig:stressprofile} shows, in ThinHR (and all our other
simulations), the fluid-frame electromagnetic stress is tens of
percent above the Novikov-Thorne prediction in the disk body, and
rises steeply inward inside $r =10M$, reaching a level $\simeq 5$ times
greater than the stress found in the Shafee et~al. calculation.
Similarly, although we found $u_\phi (\rm{ISCO}) - j_{\rm net} \simeq 0.33$,
Shafee et~al. found a value less than half as large, only $\simeq 0.14$.

The origin of this contrast is uncertain.
Although we have not yet performed a simulation with several dipolar
loops, the very minor contrast between Kerr simulations with dipolar
and quadrupolar initial field suggests that different forms of poloidal
field are not, by themselves, significant.  However, different field
structures can place different demands on spatial resolution;
as clearly demonstrated by MediumLR (and perhaps by KD0), inadequate resolution
can lead to substantial artificial suppression of magnetic field strength.
The figures illustrating results from MediumLR that we have shown in this paper
all draw on data from the period during that simulation when it remained well-resolved;
at later times, as its resolution quality failed, its electromagnetic stresses steadily
weakened and $u_\phi (\rm{ISCO}) - j_{\rm net}$ diminished.
Because it entails more small scale structure, a pair of dipole loops, as
in the Shafee et~al. initial condition, may
create a turbulent magnetic field more vulnerable to reconnection than
our single dipolar loop, particularly when perturbed by $50\%$ and
studied with a grid having larger radial cells.

It is possible that two other considerations may also play a role in
creating these contrasting conclusions.  First, their simulation ran only
to a time of $10000M$, and they presented no data demonstrating how well, and over
what range of radii, it reached a state of inflow equilibrium.  Because
we found that a disk of this thickness takes $\simeq 10000M$ to reach
equilibrium, their shorter duration may be problematic.  Second,
the Shafee et~al. simulation had an azimuthal range of only $\pi/4$.
\citet{2006ApJ...651.1031S} showed that the characteristic azimuthal
coherence length of features in full $2\pi$ global disk simulations was
$\simeq 1$~radian.  This result suggests that an azimuthal extent of
only $\pi/4$ might misrepresent the MHD turbulence.

\section{Discussion}

It is almost twenty years since the first recognition \citep{BH91} that
magnetic fields can produce significant stress with disks; it has
been ten years since the first suggestions \citep{K99,G99} that this
stress could continue within the ISCO, with implications for
the overall efficiency and luminosity of accretion disk.  However,
there remains controversy about how large these effects may be and
how they depend on disk parameters.  Because the first general relativistic
MHD accretion simulations demonstrated that these stresses are significant
in disks with thickness $H/R \simeq 0.15$ \citep{KHH05}, the discussion
in the last several years has centered on whether they might diminish
with decreasing disk thickness \citep{RF08,2008ApJ...687L..25S}.
In this paper we have carried out a series of simulations carefully
designed to isolate the effect of
varying $H/R$; we find in all of our simulations that the electromagnetic
fluid-frame stress increases steeply inward almost all the way to the
event horizon.  Indeed, even for a contrast of a factor of 3 from the thickest to
the thinnest disk, we find an almost imperceptibly small change in the
fluid-frame electromagnetic stress profile.  The most natural interpretation
of our results is that the radial distribution of electromagnetic
stress depends at most only weakly on disk thickness.  In our simulations,
these stresses diminish $j_{\rm net}$, the net angular momentum per unit
rest-mass accreted, by $\sim 10\%$.
This result supports quantitatively the very crude qualitative argument
given in \citet{K99} that the Alfv\'en speed in the plunging region
would always become marginally relativistic, more or less independent
of the accretion rate, so that magnetic stresses there would always be
significant, but would always be dominated by gravitational forces.

It is also in keeping with other lines of qualitative reasoning.
One might begin by asking, ``If stirring of MHD turbulence by the MRI
leads to significant magnetic stresses in the disk body, what might
change near the ISCO?"  As shown by \citet{GP98}, the orbital shear
in the plunging region in a Kerr spacetime differs from the Newtonian
value by only a number of order unity.  Consequently, if the magnetic
pressure continues to be smaller than the gas pressure in the disk's
equatorial plane, one would expect linear growth of the MRI to behave
in very much the same way as in the disk body.  If the gas pressure
falls relative to the magnetic pressure so that the plasma $\beta$
drops below unity, thereby quenching linear growth of the MRI, then
magnetic stresses are surely important.  However, nonlinear development
of the turbulence can be expected to change as the infall time becomes
as short or shorter than the eddy turnover time, the time for the energy
of turbulent motions to move from long length scales to short.  Indeed,
earlier work \citep{2004ApJ...606.1083H} has shown that the magnetic
field becomes both much smoother and somewhat less tightly-wound (i.e.,
the radial component grows somewhat relative to the toroidal component)
just inside the ISCO.  Such a change in structure could alter both the
rate of field amplification and the rate of field dissipation.  On the
basis of simple arguments like these, however, it is difficult to say
whether these changes should lead to a larger or smaller mean field
intensity and therefore stress.  In the limit that both amplification
and dissipation become weaker, flux-freezing results in growing magnetic
stresses \citep{K99,G99}.  Moreover, with the principal dynamics---orbital
shear and a growing radial velocity---all acting in the equatorial plane,
there is no obvious place for a dependence on disk thickness.

This physical argument is bolstered by the fact that the different
simulation versions differ only slightly.  Significantly different
grid schemes, contrasting initial conditions, and even wholly different
codes make only slight differences in the outcome.  Even the topological
contrast of substantial net vertical flux versus none at all seems to
change the stress by only a modest amount.

On the other hand, we have also found that fluid effects in the inner
disk can also contribute to a diminution in $j_{\rm net}$.
Pressure support of the matter in the disk is, by definition,
proportional to $(H/R)^2$.  Consequently, the mass-weighted mean angular
momentum at any location in the disk is smaller than the angular
momentum of a test-particle at that radius by a comparable amount.
More significantly, Reynolds stresses can reduce it further, by an amount
that increases both with increasing disk thickness and with net magnetic
flux trapped on the horizon.

A few notes of caution should be injected into this discussion, however:
Previous work studying accretion in a Kerr ($a/M = 0.9$) geometry
\citep{BHK08} suggests that a disk magnetic field that has no net poloidal
content might produce weaker electromagnetic stress both throughout the
disk and in the ISCO region.  This effect should be explored more thoroughly
in future work.  We have also shown that before interpreting simulation
results in terms of their implications for steady-state accretion, it
is important to check carefully both that the simulation approximates
inflow equilibrium and that the simulation's spatial grid provides
adequate resolution throughout the period studied.

We conclude, then, that there appears to be little evidence for a
strong dependence of near-ISCO electromagnetic stress on either disk
thickness or the net magnetic flux.  Because disk thickness is really
a function of accretion rate, substantial near-ISCO electromagnetic stress
should be seen in black hole
accretion systems whether they are accreting at a rate near Eddington
or far below.  Its weak dependence on net magnetic flux suggests that
the impact of electromagnetic stresses should be similarly independent
of external magnetic boundary conditions.  At the same time, we also
find a supplemental reduction of the net accreted angular momentum
due to Reynolds stresses, and this depends on both $H/R$ and
magnetic geometry.  When the Reynolds stress is weakest, so only the
near-universal electromagnetic stress acts, $j_{\rm net}$ is reduced
below $u_\phi (\rm{ISCO})$ by $\simeq 7$--$10\%$; when the Reynolds stress
is strongest, the reduction is as large as $15\%$.

Fully quantitative conclusions, however, await several
extensions of this work: to rotating black holes, to disks with more
complex magnetic topologies, and to disks whose scale heights are constant,
rather than proportional to radius.  In the radiation-dominated regime
(which should apply to the inner regions of accretion disks around black
holes whenever the accretion rate is near Eddington, particularly when
the central mass is large: \cite{SS73}), $H$ is constant as a function
of radius.  Therefore, a flat-topped disk is a more realistic model for
disks as we are likely to find them in Nature.  All of these extensions
should be feasible in the near-term.

\acknowledgments{ This work was supported by NSF grant AST-0908869 and
NASA grant NNX09AD14G (JFH), and by NSF grant AST-0507455 (JHK).
Some of the simulations described were carried out on the Teragrid Ranger
system at TACC and the Teragrid Abe system at NCSA, both supported by 
the National Science Foundation.  Other simulations were run on
the Johns Hopkins Homewood High-Performance Computing Center cluster.}

\bibliography{bib}

\clearpage

\clearpage

\clearpage

\clearpage

\clearpage

\clearpage

\clearpage

\clearpage

\clearpage

\clearpage

\clearpage

\clearpage

\clearpage
\begin{figure}
\centerline{\includegraphics[angle=90,scale=0.4]{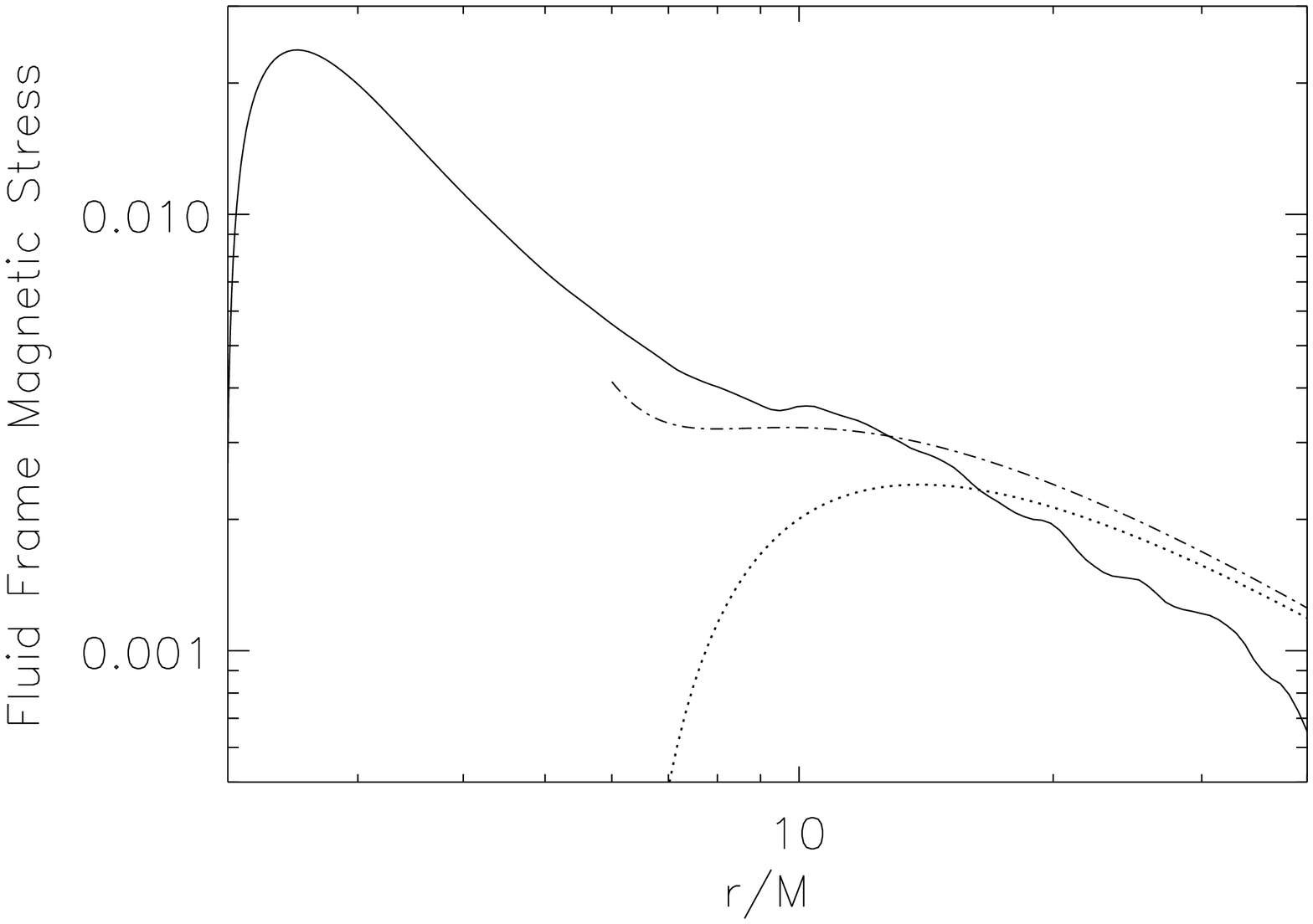}
\includegraphics[angle=90,scale=0.4]{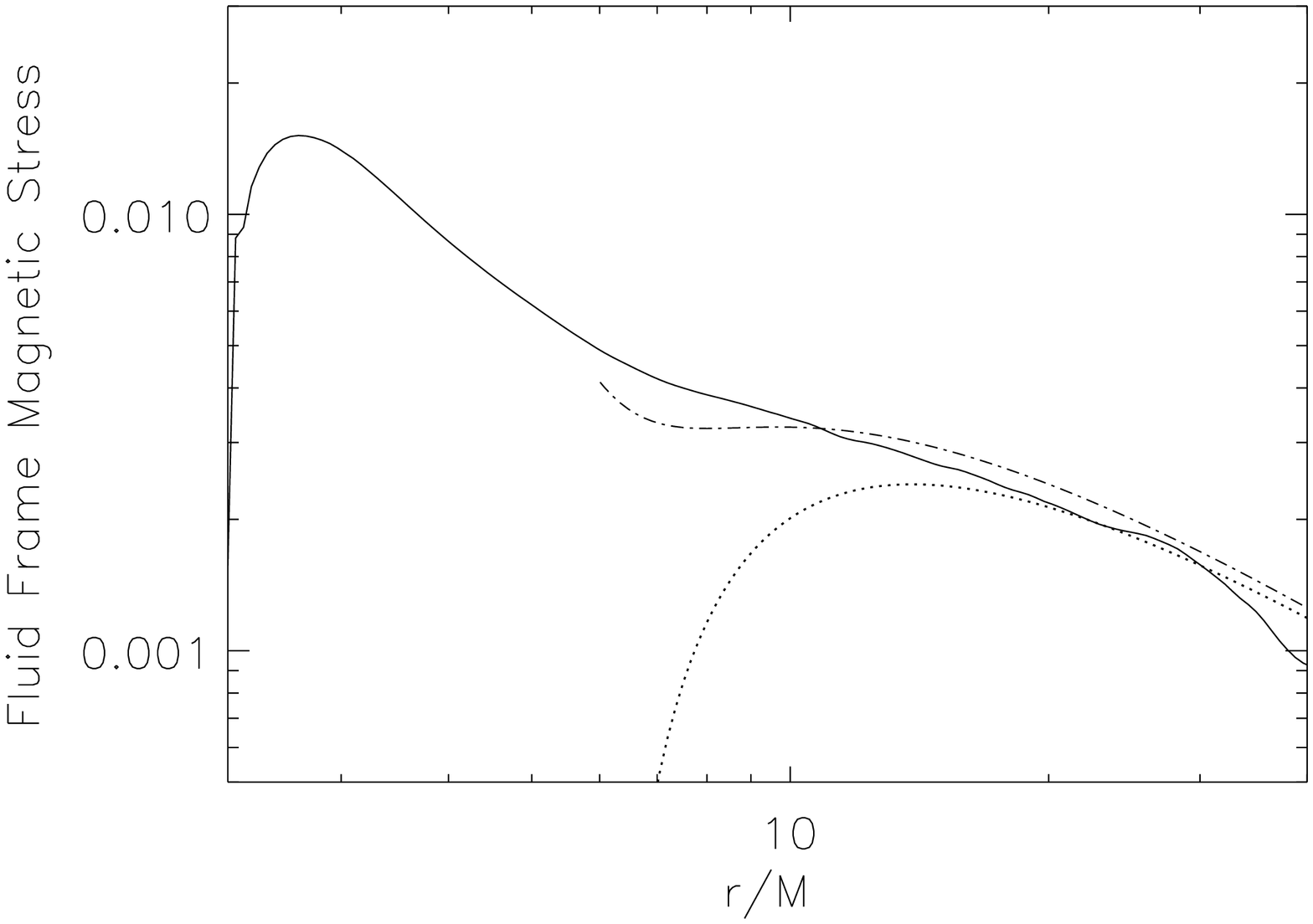}}
\caption{Fluid-frame electromagnetic stress, normalized by that 
simulation's mean
accretion rate, for the GRMHD simulations.  (Left) KD0c.  (Right) VD0.
Curve identifications are as in Fig.~\ref{fig:stressprofile}.
\label{fig:vertstressprofile}}
\end{figure}

\clearpage
\begin{figure}
\centerline{\includegraphics[scale=0.4]{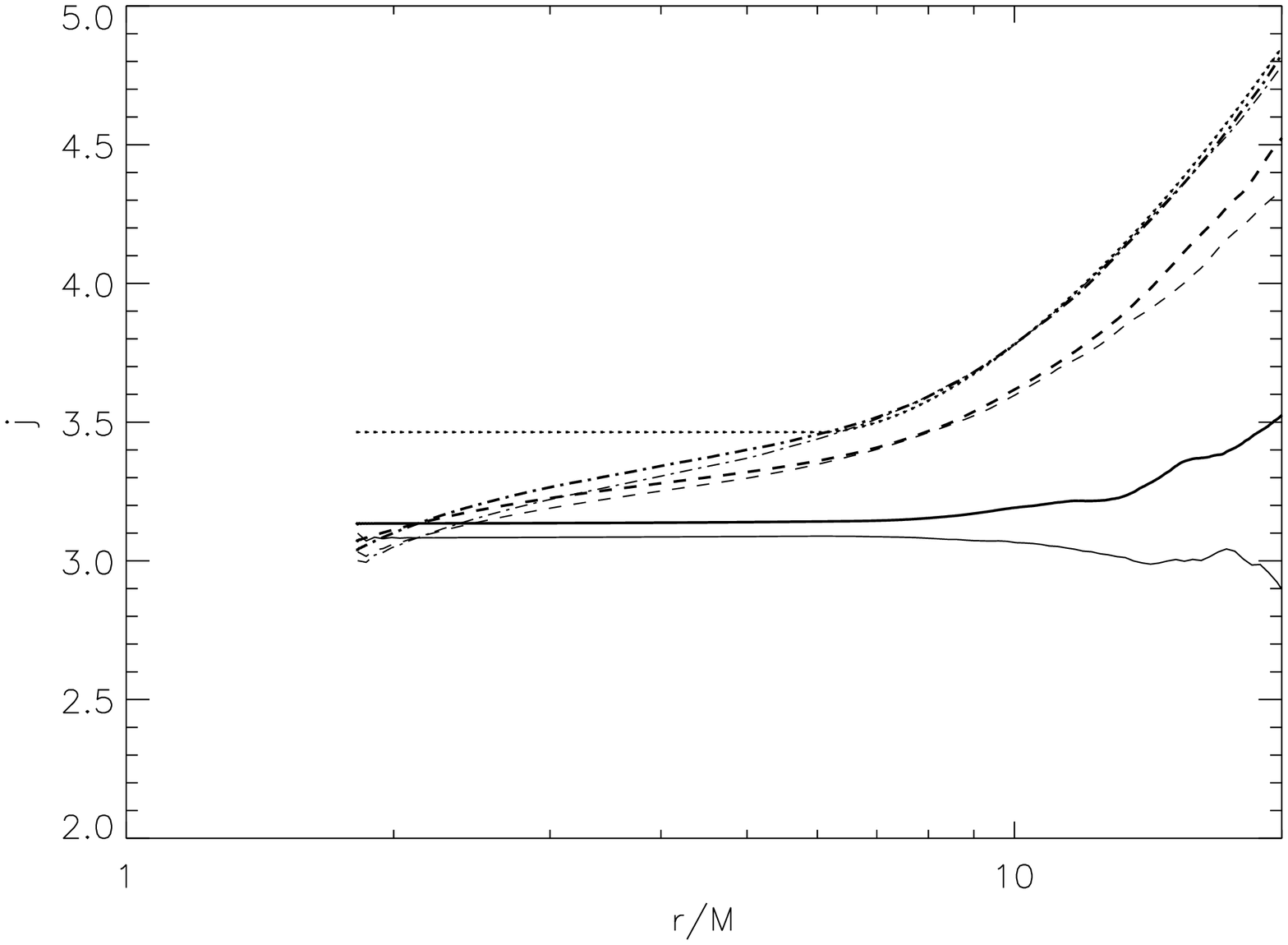}
\includegraphics[scale=0.4]{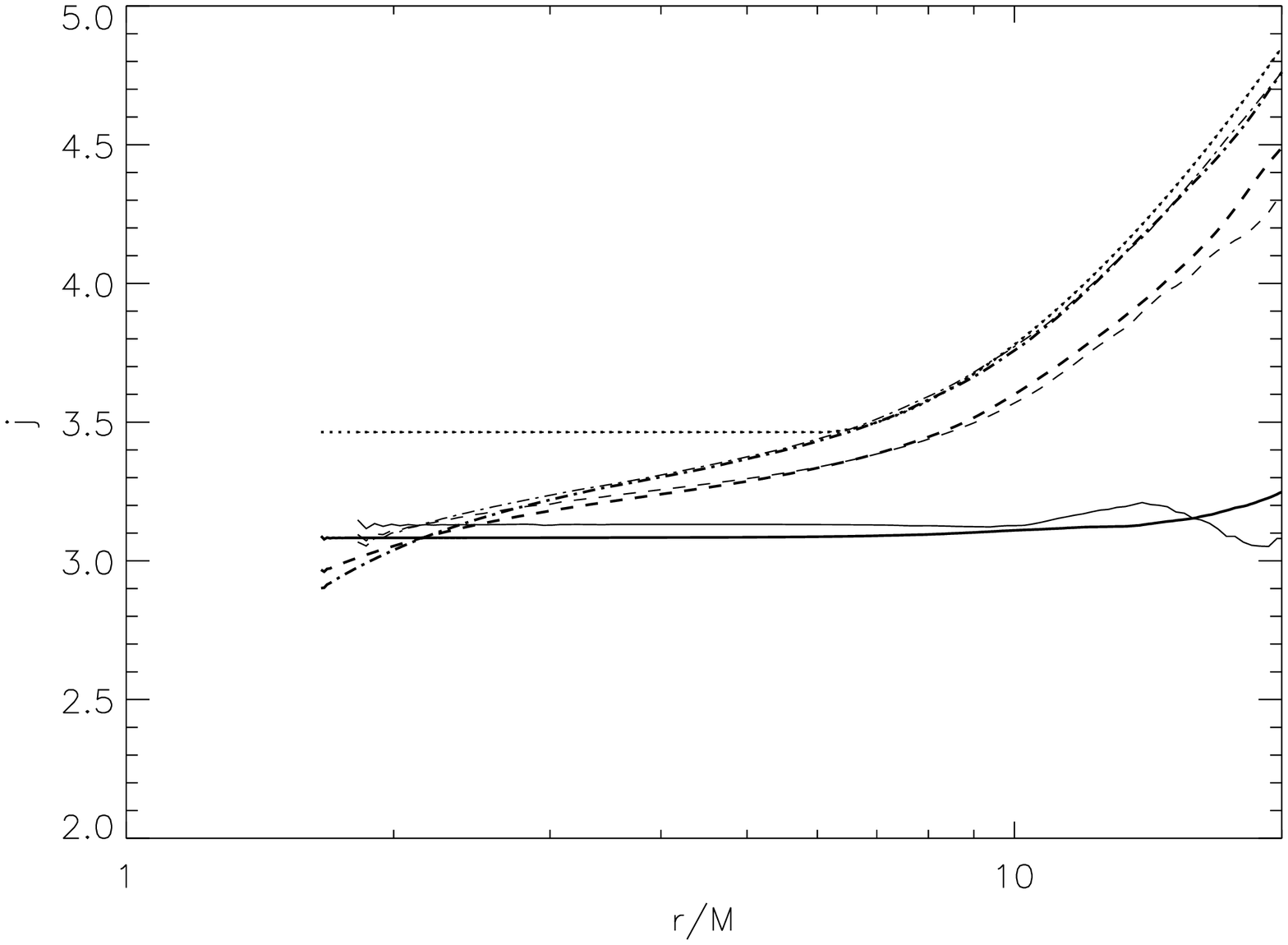}}
\centerline{\includegraphics[scale=0.4]{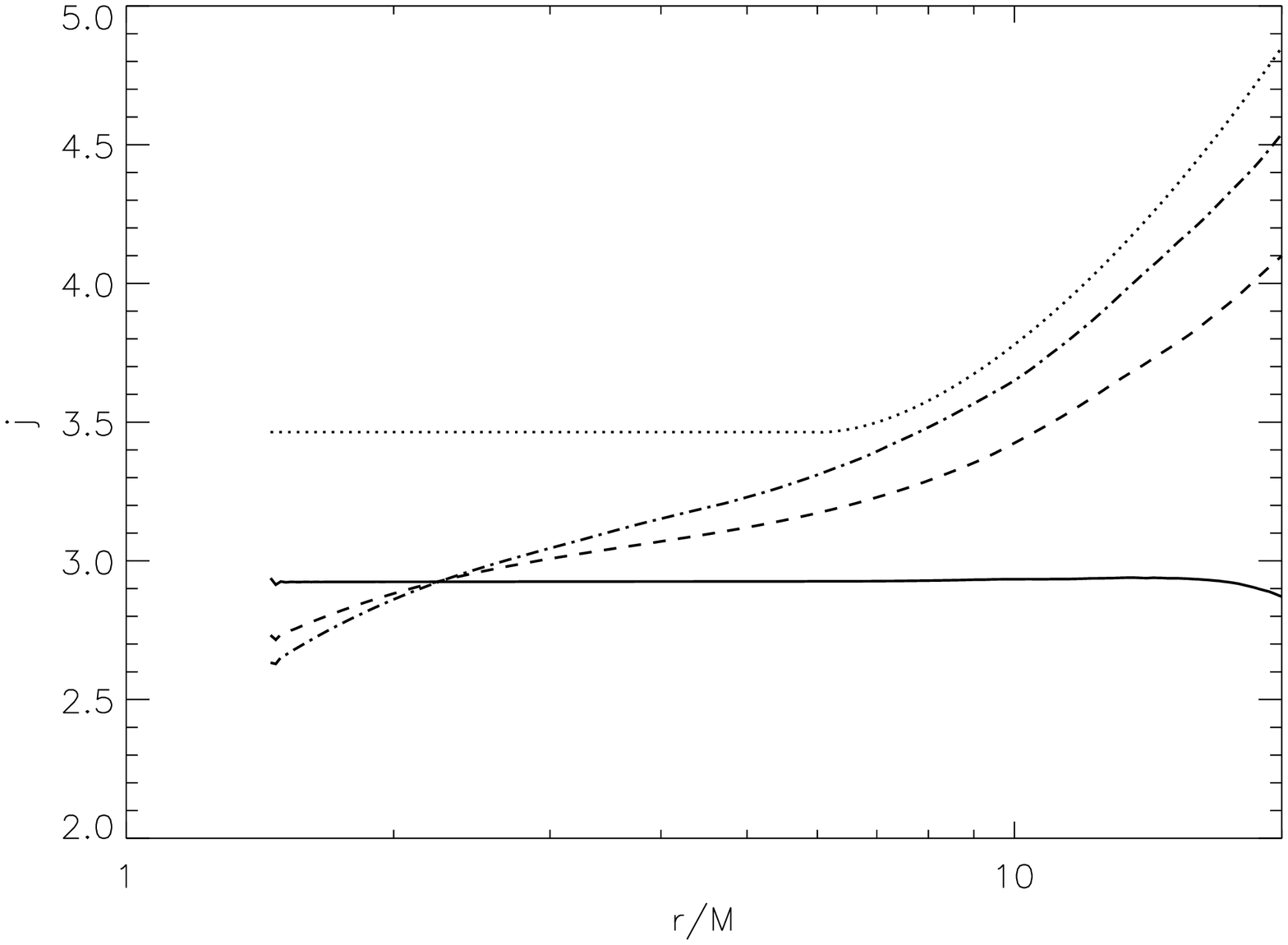}}
\caption{
Accreted angular momentum per unit rest-mass for each of the three
aspect ratios simulated with HARM3D.  (Top left) ThinHR and ThinLR, 
(top right) MediumHR and MediumLR, (bottom) ThickHR.  Solid curves
show $j_{\rm net}$, with the heavy curves corresponding
to HR, the light curves to LR.  Dotted curves show the
angular momentum of a circular orbit as a function of radius; inside the ISCO,
it is held constant at $u_\phi(\rm{ISCO})$, consistent with the
Novikov-Thorne model.  
Dashed curves represent the time-averaged
specific angular momentum carried by the accreting matter, i.e., $\langle
\rho h u^r u_\phi\rangle/\langle \rho u^r \rangle$.  Dot-dashed curves
are the time-averaged mass-weighted mean angular momentum at each radius,
i.e., $\langle \rho h u_\phi \rangle/\langle \rho \rangle$.
\label{fig:netangmom}}
\end{figure}

\clearpage
\begin{figure}
\centerline{\includegraphics[angle=90,scale=0.4]{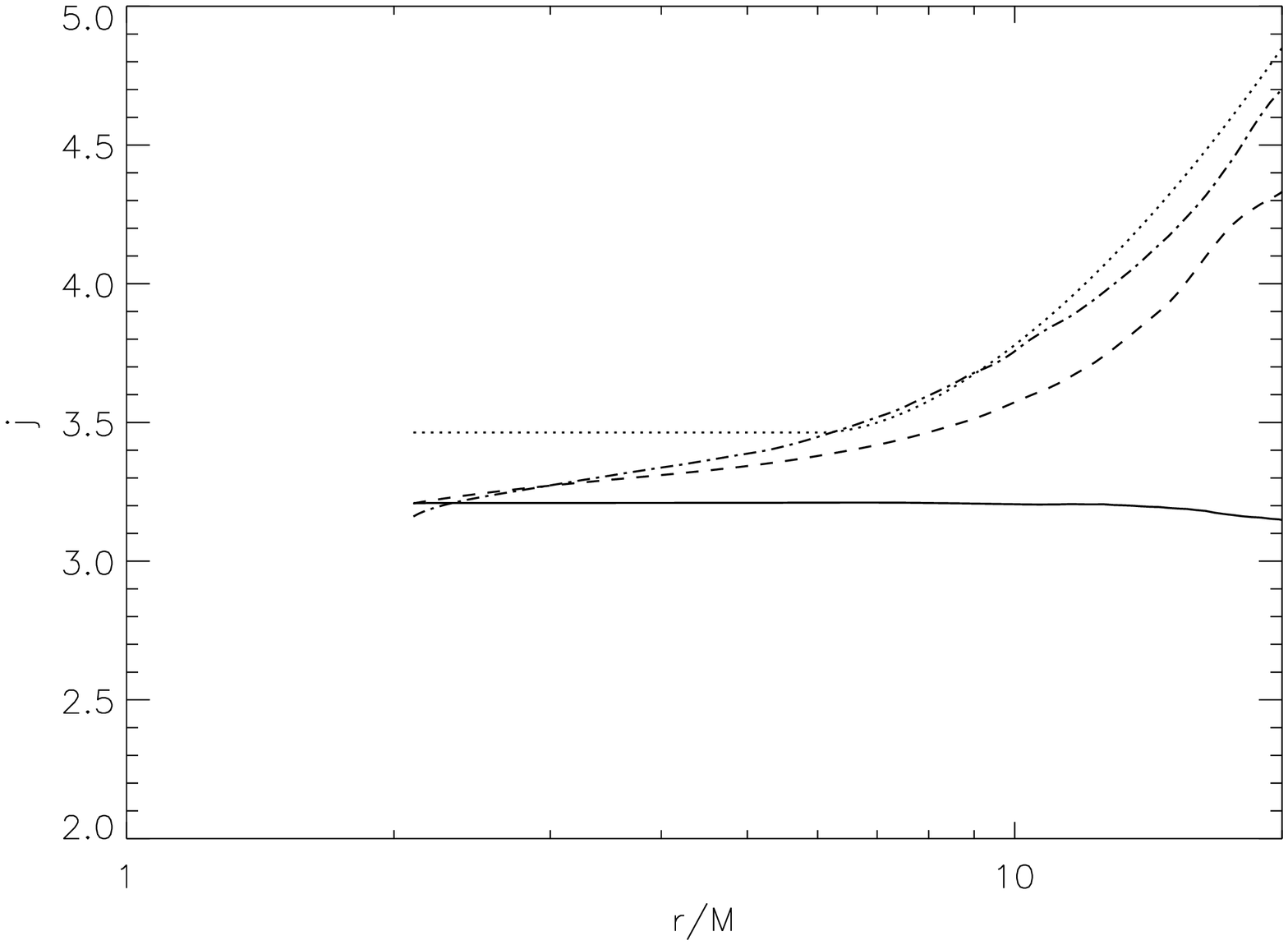}
\includegraphics[angle=90,scale=0.4]{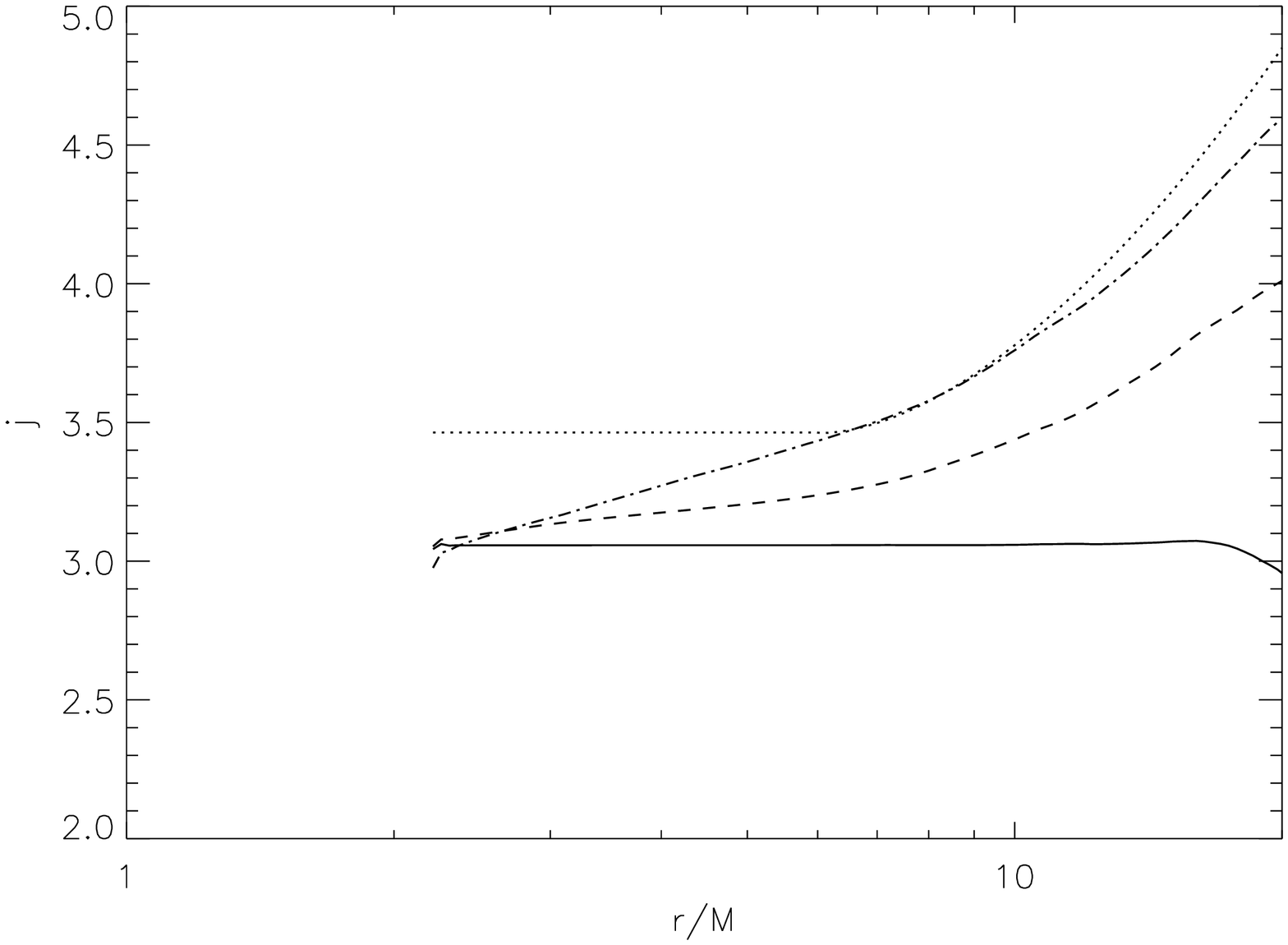}}
\caption{
Accreted angular momentum per unit rest-mass for the two GRMHD
simulations.  (Left) Simulation KD0c.  (Right) Simulation VD0.
Curves have the same identifications as in Fig.~\ref{fig:netangmom}.
\label{fig:KD0VD0netangmom}}
\end{figure}

\clearpage

 \begin{deluxetable}{lcccccccccc}
 \tablecolumns{11}
 \tablewidth{0pc}
 \tabletypesize{\footnotesize}
 \rotate
 \tablecaption{Simulation Parameters}
 \tablehead{
 \colhead{Name}         &
 \colhead{\begin{tabular}{c} Target \\ $H/R$ \end{tabular}} &
 \colhead{Cell Count\tablenotemark{a}}         &
 \colhead{$\theta$ Grid}         &
\colhead{$\xi$}         &
 \colhead{$\theta_c$}         &
\colhead{$\theta_0$}         &
\colhead{$s$}         &
 \colhead{\begin{tabular}{c} Cell Shape\tablenotemark{b} \\ ($\theta=\pi/2$) \end{tabular}} &
 \colhead{\begin{tabular}{c} Cell Shape\tablenotemark{b,c} \\ ($\theta=2H/R$) \end{tabular}} &
 \colhead{\begin{tabular}{c}$\Delta\theta_\mathrm{min}/10^{-3}$ \\ \end{tabular}} 
 }
 \startdata
 ThinHR   &  0.05              & $912\times 160 \times 64$ &(\ref{theta-x2-2-def}) &$0.93$  &$\pi\cdot10^{-15}$ &$\ldots$   &$\ldots$    &$3.0:1:17$  &$0.48:1:2.7$ &$1.4$  \\
 ThinLR   &  0.05              & $192\times 192 \times 64$ &(\ref{theta-x2-1-def}) &$0.49$  &$10^{-15}$         &$0$        &$\pi$           &$71:1:74$   &$2.0:1:2.1$  &$0.33$ \\
 MediumHR &  0.08              & $512\times 160 \times 64$ &(\ref{theta-x2-2-def}) &$0.93$  &$\pi\cdot10^{-15}$ &$\ldots$   &$\ldots$      &$5.8:1:17$  &$0.43:1:1.3$ &$1.4$  \\
 MediumLR &  0.08              & $192\times 192 \times 64$ &(\ref{theta-x2-1-def}) &$0.35$  &$0.083$            &$\theta_c$ &$0.95$          &$5.1:1:5.3$ &$2.7:1:2.7$  &$4.7$  \\
 ThickHR  &  0.16              & $348\times 160 \times 64$ &(\ref{theta-x2-2-def}) &$0.76$  &$0.094$            &$\ldots$     &$\ldots$    &$3.0:1:5.3$ &$0.87:1:1.4$ &$4.6$  \\
 KD0c     & $\ldots$\tablenotemark{d} & $192\times 192 \times 64$ &\citep{HK06}          &       &$0.045\pi$             &     &             &$2.4:1:2.3$ &$2.1:1:1.9$  &$11$   \\
 VD0      & $\ldots$\tablenotemark{d} & $256\times 256 \times 64$ &\citep{BHK09}          &       &$0.01\pi$       &        &     &$2.7:1:2.9$ &$2.3:1:2.4$  &$8.4$  \\
 \enddata
 \label{tab:simdefs}
\tablenotetext{a}{$N_r \times N_\theta \times N_\phi$}
\tablenotetext{b}{$\sqrt{g_{rr}} \Delta r : \sqrt{g_{\theta\theta}} \Delta \theta : \sqrt{g_{\phi\phi}} \Delta \phi$}
\tablenotetext{c}{Values of $H/R$ used here are the actual time-averaged 
quantities specified in Table~\ref{tab:simresults}}
\tablenotetext{d}{Unregulated scale height.}
 \end{deluxetable}

\clearpage

\begin{deluxetable}{lcccc}
\tablecolumns{5}
\tablewidth{0pc}
\tablecaption{Simulation Results}
\tablehead{
\colhead{Simulation}         &
\colhead{Actual $\langle H/R \rangle$\tablenotemark{a}} &
\colhead{$j_{\rm net}$} &
 \colhead{$N_{\rm cells}(|z|<H/R)$\tablenotemark{c} } & 
 \colhead{$\Delta t_{\rm ave} / \left(10^3 M\right)$}
}
\startdata
ThinHR   &$0.061$                 & 3.13   &$81$   & $10$--$15$    \\
ThinLR   &$0.085$                 & 3.07   &$60$   & $4.5$--$11.5$ \\
MediumHR &$0.10$                  & 3.08   &$103$  & $5$--$12.5$   \\
MediumLR &$0.091$                 & 3.10   &$35$   & $5$--$8$      \\
ThickHR  &$0.17$                  & 2.93   &$74$   & $8$--$13.66$  \\
KD0c    &$0.13$\tablenotemark{b} & 3.21   &$24$   & $4$--$10$     \\
VD0     &$0.14$                  & 3.06   &$32$   & $14$--$20$    \\
\enddata
\label{tab:simresults}
\tablenotetext{a}{Average of $H(r,t)/R$ 
in time (over each run's $\Delta t_{\rm ave}$) and in radius (weighted uniformly 
with respect to $\log r$) from $\risco$ to $\rpmax$.}
\tablenotetext{b}{Averaging period used here was $8000M-10000M$ instead of
$\Delta t_{\rm ave}$ because only these data were saved.}
\tablenotetext{c}{Number of cells per scale height averaged with respect to $\log r$,
i.e., $\int \mathcal{N}(r) \, d\log\! r / \int d\log\! r$, where 
$\mathcal{N}(r)$ is the number of cell widths within 
$\theta \in \left[\pi/2 - H(r)/R \, , \, \pi/2 + H(r)/R\right]$ and the bounds 
of integration are $r\in\left[\rhor , \rpmax\right]$.  }
\end{deluxetable}

\end{document}